\documentclass[12pt,apj]{emulateapj}
\slugcomment{{\sc Accepted to ApJ:} May 06, 2008}

\shorttitle{Activity, Rotation, and Braking in M and L Dwarfs}

\begin{document}


\title{Chromospheric Activity, Rotation, and Rotational Braking in
 M and L Dwarfs}


\author{A. Reiners\altaffilmark{*}} 
\affil{Institut f\"ur Astrophysik, Georg-August-Universit\"at, D-37077
  G\"ottingen, Germany}
\email{Ansgar.Reiners@phys.uni-goettingen.de}
\altaffiltext{*}{Emmy Noether Fellow}

\and

\author{G. Basri}
\affil{Astronomy Department, University of California, Berkeley, CA
  94720 }
\email{basri@berkeley.edu}



\begin{abstract}
  We present results from a high-resolution spectroscopic survey of 45
  L dwarfs, which includes both very low-mass stars and brown dwarfs.
  Our spectra allow us to derive a significant number of new
  rotational velocities, and discover a slowly rotating (in projected
  velocity) L dwarf that allows more accurate measurement of
  spectroscopic rotations for these objects. We measure chromospheric
  activity (and often its variability) through the H$\alpha$ emission
  line. Our primary new result is good evidence that magnetic braking
  dominates the angular momentum evolution of even brown dwarfs,
  although spindown times appear to increase as mass decreases. We
  confirm that activity decreases as effective temperature decreases,
  though a larger fraction of L dwarfs are active than has previously
  been reported. Essentially all active objects are also variable.  We
  confirm the lack of a rotation-activity connection for L dwarfs. We
  find a minimum limit for rotational velocities that increases with
  later spectral types, rising from near zero in older mid-M stars to
  more than 20 km s$^{-1}$ for mid-L objects. There is strong evidence
  that all L dwarfs are rapid rotators. We derive a braking law that
  can depend on either temperature or mass which can explain all the
  rotational results and provides an age dependence for the angular
  momentum evolution. It is clear that angular momentum loss
  mechanisms in smaller and cooler objects become more inefficient,
  starting at the fully convective boundary.
\end{abstract}






\section{Introduction}

The evolution of magnetic activity and angular momentum in low-mass
stars and brown dwarfs provides fundamental insights into the nature
of low-mass objects and fully convective dynamos.  Rotation and
activity in low-mass stars and brown dwarfs are discussed in the work
by, for example, \citet{Mohanty03, Liebert03, West04, Zapatero06,
  Schmidt07}. In comparison to early-M stars, normalized activity
$\log{L_{\rm{H_{\alpha}}}/L_{\rm{bol}}}$ does not weaken until
spectral type mid-M; after that it diminishes, but flares are still
observed in early L dwarfs. Among the M dwarfs, \citet{Mohanty03}
found evidence for a rotation-activity connection down to spectral
type M8.5 but a sharp break in this connection in cooler objects than
that.

The investigation of rotational evolution of low-mass stars and brown
dwarfs is complicated by several effects.  At young ages, they all
contract gravitionally, leading to rotational spin-up. Thus objects of
ages between 10-100\,Myrs generally are rapidly rotating
\citep[e.g.][]{Barnes03}. On the ZAMS, stars do not significantly
contract any further but instead suffer rotational braking through a
magneto-thermal wind that is driven by magnetic activity
\citep[e.g.][]{Mestel99}. Among low-mass stars, this braking is
strongest in early-M stars among which almost all have rotational
velocities below the detection limit \citep{Marcy92, Delfosse98}.
Brown dwarfs don't reach the ZAMS, but do finish most of their
contraction after about a Gyr.

A very large fraction of early M stars are rotating too slowly to be
detected by spectroscopic broadening. A sudden change in rotational
braking appears around spectral type M3, in the sense that a large
fraction of stars later than M3 are rapid rotators compared with early
M stars. The rapidly rotating mid-M dwarfs are members of the young
disk population while older mid-M stars tend to rotate more slowly
\citep{Delfosse98}. This means that rotational braking is still
working in mid-M type stars, but that spindown times are longer
than for the early-M dwarfs. An age-activity relation has recently
been quantified by \citet{West08}, which could be related to rotational
braking and a rotation-activity connection. The difference in rotational
braking between early- and mid-M stars is possibly due to the fully
convective nature of the latter -- the threshold at which rapid
rotation occurs coincides with the threshold to complete convection.
If so, it is more likely a manifestation of field geometry than field
strength, since fully convective stars produce fields at least as
strong as partially convective stars \citep{RB07}.

Objects of spectral type M6.5 or later can either be stars at ages of
several Gyrs, or they can be young brown dwarfs. As they age, brown
dwarfs leave the M spectral range, evolving towards lower temperatures
through the spectral types L and/or T \citep{Burrows97, Baraffe03}.
During this evolution, they continuously contract, leading to constant
rotational acceleration. The rotational evolution of brown dwarfs is
not well known. Rotational velocities of young brown dwarfs were
measured by \citet{Mohanty05}.  The results from measuring rotational
periods in young brown dwarfs are summarized in \cite{Herbst07}.
Rotation in field late-M and L dwarfs was measured by
\citet{Mohanty03}, and in objects down to spectral class T by
\citet{Zapatero06}. In the latter work the authors provide a model for
the rotational evolution of brown dwarfs based on gravitational
contraction. This model explains the lack of slowly rotating late-L
dwarfs by contraction during the lifetime of a brown dwarf.

In this paper, we provide measurements of rotation and chromospheric
(H$\alpha$) activity in a substantial sample of 45 L dwarfs. We
investigate activity among low-mass stars and brown dwarfs. We measure
rotational velocities for our sample.  Together with velocities taken
from the literature, we can investigate a comprehensive sample of
ultra-cool stars and brown dwarfs with measured rotation velocites. We
use this sample to constrain the rotational evolution of brown dwarfs,
and show that the present data indicates that wind-braking is still
important in brown dwarfs.

\section{Observations and Data Sample}

To investigate rotation and activity in L dwarfs, we collected a large
sample of high resolution L dwarf spectra. This sample consists of new
spectra we took at HIRES/Keck and UVES/VLT, and of spectra taken from
the UVES data archive which we reanalyze. We include in our analysis
results on rotation and H$\alpha$ emission adopted from the
literature. Sample selection is based on brightness criteria to
maximize the number of observed L dwarfs. This means that the sample
is very heterogeneous and probably biased towards young (hence bright)
targets. Sample data are shown in a spectral atlas covering the
wavelength range 6400--10200\,\AA\ in \cite{R07b}.

\subsection{New Observations}

The first part of our data set consists of 61 high resolution spectra
of 45 L dwarfs; their exposure times and SNR as measured in the FeH
band are shown in Table\,\ref{tab:observations}. All spectra cover the
spectral range from H$\alpha$ up to 1\,$\mu$m, including the Wing-Ford
band of molecular FeH. 22 spectra were obtained by us with the HIRES
spectrograph at Keck observatory.  The three HIRES CCDs cover the
spectral region from 5700\,\AA\ to 10\,000\,\AA\ in one exposure.
HIRES spectra were taken at a slit width of 1.15\,\arcsec, yielding a
resolving power of $R \sim 31\,000$. In addition to that, 39 spectra
were taken with the UVES spectrograph at the VLT. 24 of them were
observed in service mode for our program (PIDs 077.C-0449 and
078.C-0025); the remaining 15 are from the ESO science archive under
PID 065.L-0199 \citep[PI C.A.L. Bailer-Jones, see][]{Bailer04}. The
UVES spectra from our program and the ones from the archive are taken
with similar settings.  Our new data are centered at 830\,nm at a slit
width of 1.2\,'' ($R \sim 33\,000$), archive data are centered at
837\,nm with a slit width of 1.0\,'' ($R \sim 40\,000$). Both setups
cover the wavelength region 6400--10\,250\,\AA\ on two CCDs.

\subsection{Data from the Literature}

In our discussion of rotation and activity we enrich our sample with
three high spectral resolution samples from the literature.
\cite{Mohanty03} give rotation velocities and H$\alpha$ equivalent
widths of mid- to late-M dwarfs and 13 L dwarfs. Their data were taken
with the old HIRES CCD before the detector was upgraded to a 3-CCD
array, hence the data does not cover the FeH band.  \cite{Delfosse98}
measured rotation and H$\alpha$ activity in a large sample of
M-dwarfs. We use their results to compare activity between M and L
dwarfs. \cite{Mohanty03} and \cite{Delfosse98} measured the projected
rotation velocities with the cross-correlation technique using a
slowly rotating mid-M star as a template. This method can be expected
to be very accurate among the M dwarfs. For the L dwarfs, however, we
suggest below that such a measurement can be affected by the mismatch
of the template's and the target's spectral types.

\cite{RB07} report rotation velocities, H$\alpha$ activity and the
magnetic flux for 22 M dwarfs. In the work presented here, we follow
the same analysis strategies but without the measurement of magnetic
flux, which becomes quite difficult for the L dwarfs in their heavily
saturated FeH band. We include the sample from \cite{RB07} in the comparison
of the M dwarf and L dwarf rotation-activity connection.

\section{Analysis}

\subsection{Stars with Li-absorption}

Our spectral coverage includes the Li line at $\lambda = 6708$\,\AA.
We find Li absorption in three objects, 2MASS0602$-$39 (L1.0,
$v\,\sin{i} = 9$\,km\,s$^{-1}$, no H$\alpha$ emission), 2MASS2057$-$02
(L1.5, $v\,\sin{i} = 62$\,km\,s$^{-1}$,
$\log{L_{\rm{H_{\alpha}}}/L_{\rm{bol}}} = -4.9$), and we confirm its
existence in 2MASS1305$-$25 (Kelu-1, L2.0, $v\,\sin{i} =
76$\,km\,s$^{-1}$). At such low temperature, this means that the three
objects are substellar and likely younger than a few
hundred million years \citep{Basri00}.  2MASS2057$-$02 has a very low
tangential velocity of $v_{\rm tan} = 6$\,km\,s$^{-1}$
\citep{Schmidt07} consistent with youth \citep{Chiba00}.  We see no Li
absorption in any other target, among them 2MASS0523$-$14, a mid-type
L-dwarf (L2.5) with very low tangential velocity ($v_{\rm tan} =
5$\,km\,s$^{-1}$), hence probably young. Lack of lithium is not firm
evidence of a stellar nature; brown dwarfs above 60 jupiter masses will
also deplete their lithium within a few hundred million years.

\subsection{H$\alpha$ Emission}

\begin{figure}
  \plotone{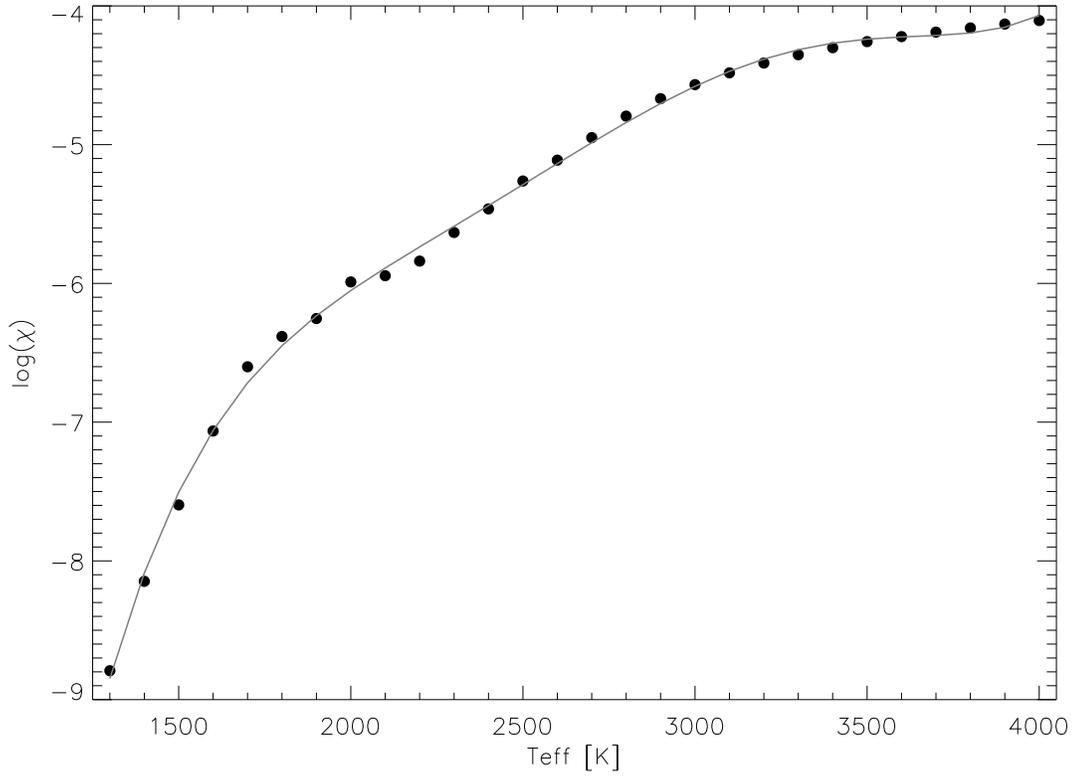}
  \caption{\label{fig:chifactor}$\log{\chi}$, with $\chi$ the specific
    H$\alpha$ luminosity normalized to the bolometric luminosity per
    \AA ngstrom equivalent width calculated from PHOENIX model
    spectra.  The grey line shows the fifth order polynomial in
    Eq.\,\ref{eq:chifactor} and Table\,\ref{tab:chifactor}.}
\end{figure}

To measure the equivalent width in the H$\alpha$ line against the
continuum, we normalize the line at two footpoints blue- and redward
of H$\alpha$. The footpoints are the median values at 6545 --
6559\,\AA\ on the left hand side, and at 6567 -- 6580\,\AA\ on the
right hand side of the H$\alpha$ line.  None of the emission lines
found in our targets extends into the region used for normalization.
The H$\alpha$ equivalent width is then measured by integrating the
flux from 6552 to 6572\,\AA. The uncertainty in our equivalent width
measurement is approximately 0.2\AA\ \citep[see][]{Mohanty03}; we
provide individual limits in Table\,\ref{tab:results}. We convert the
measured H$\alpha$ equivalent width into H$\alpha$ flux by measuring
the flux per unit equivalent width from the continuum flux in
synthetic PHOENIX spectra \citep[][we used the DUSTY
models]{Allard01}. This is the same as using a ``$\chi$ factor'' as
done by \cite{Walkowicz04} to translate equivalent widths into $F_{\rm
  H{\alpha}}/F_{\rm bol} = L_{\rm H{\alpha}}/L_{\rm bol}$ according to
$L_{\rm H{\alpha}}/L_{\rm bol} = \chi\cdot{\rm EqW(H\alpha)}$.
\cite{Walkowicz04} used observed spectra to calibrate $\chi$ and
expressed it in terms of colors. Here, we use effective temperature
from the models and measure the value of $\chi$ between 1200 and
4000\,K. To find the temperature of our targets, we used the spectral
types and calculated $T_{\rm{eff}}$ according to the conversion given
in \cite{Golimowski04}. Our calibration of $\chi$ is plotted in
Fig.\,\ref{fig:chifactor} together with a fifth order polynomial fit
(grey line)
\begin{equation}
  \label{eq:chifactor}
  \log{\chi} = a + b T_{\rm{eff}} + c T_{\rm{eff}}^2 + d T_{\rm{eff}}^3 + e T_{\rm{eff}}^4 + f T_{\rm{eff}}^5.
\end{equation}
The parameters of the fit are shown in Table\,\ref{tab:chifactor}.
The equivalent widths and the values of $L_{\rm H{\alpha}}/L_{\rm
  bol}$ for our sample L dwarfs are given in the last two columns of
Table\,\ref{tab:results}.

For a discussion of some considerations in using PHOENIX models to
calculate H$\alpha$ fluxes, we refer to \citet{Mohanty03}. For the
stars from the sample of \cite{Mohanty03}, we took their equivalent
widths and used our calibration to re-calculate the values of $L_{\rm
  H{\alpha}}/L_{\rm bol}$ for consistency.

\subsection{Rotation Velocity}

For the analysis of rotation velocities in our L dwarf sample, we
follow the same strategy as in \cite{RB07}. We adjust the absorption
depth in a template spectrum according to a curve-of-growth scaling 
procedure, and then artificially broaden this
spectrum, searching for the best fit to the target spectrum. We employ
the absorption band of FeH close to 1\,$\mu$m for the fitting
procedure. As shown in \cite{RB06}, this region is particularly rich
in structure in ultra-cool stars and brown dwarfs, while it is also
virtually free of telluric absorption.  The absorption band of
molecular FeH is not pressure-broadened and it is embedded in a
relatively clear continuum, at least in the late M and early L dwarfs.
This is a marked difference to most other spectral regions in M- and
L-dwarf spectra, since most of their absorption features are either
poor in structure (e.g., dense molecular bands of TiO and VO) or
heavily pressure broadened (e.g., lines of alkali atoms), so that slow
rotation is difficult to detect in them.

In \cite{RB07}, we used mid-M stars as templates to fit the FeH bands
searching for rotational broadening. The depth of the absorption band
of the M dwarf spectra was scaled using a curve-of-growth scaling
\citep{RB06}. With this approach we do not need a slowly rotating
template star of each spectral type, rather we require only one slowly-
rotating template star. As a very first step, we applied the same
fitting procedure using an M dwarf template to our L dwarf spectra. We
found that the overall fit quality is still quite acceptable, and that
the scaled M dwarf template already provides very useful results
for the values of $v\,\sin{i}$ among the L dwarfs. However, the FeH
band is already somewhat saturated at spectral type L0, and the
scaling procedure becomes sensitive to the right choice of the maximum
absorption depth and the placement of the continuum in the M dwarf
spectrum. Thus, it is much better to use a spectrum of an L
dwarf as a template instead of scaling the spectrum of an M dwarf
showing much weaker FeH absorption.

\begin{figure}
  \plotone{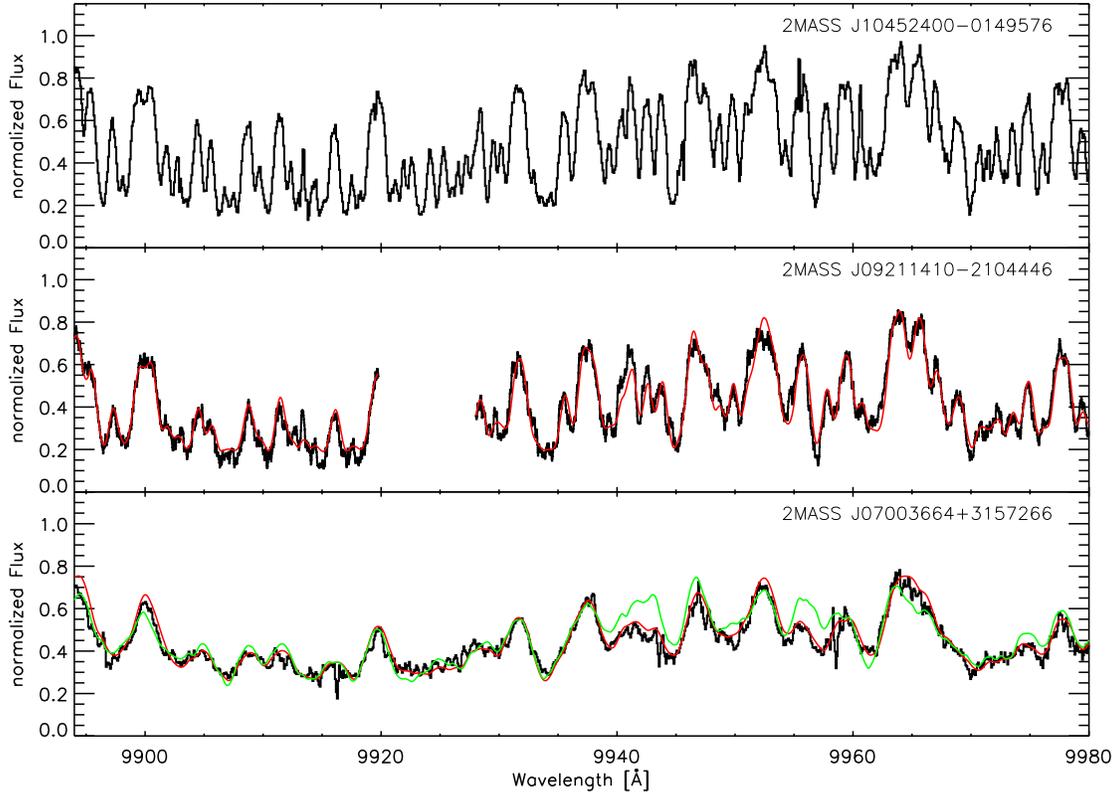}
  \caption{\label{fig:spectra}Spectra of three L dwarfs in the FeH
    molecular band. Top: HIRES-spectrum of the slow rotator 2MASS~1045
    (L1) that we used as template for rotation velocity measurements.
    Center: UVES spectrum of 2MASS~0921 (L1, $v\,\sin{i} =
    15$\,km\,s$^{-1}$); the UVES spectra have an inter-order gap at
    9920--9930\,\AA. In red the scaled and artificially broadenened
    spectrum of 2MASS~1045 is overplotted. Bottom: Same as central
    panel but for the HIRES spectrum of 2MASS~0700 (L3.5, $v\,\sin{i}
    = 41$\,km\,s$^{-1}$). In green, a scaled and broadened version of
    the spectrum of the M4.5 dwarf GJ~1227 is also shown.}
\end{figure}

In our first attempt to derive projected rotation velocities using
the M dwarf template, we found that among the 45 L
dwarfs of our sample, one object showed much narrower lines than all
the others. The L1.0 dwarf 2MASS1045$-$01 does not exhibit any sign of
extra broadening when compared to a scaled spectrum of a slowly
rotating M dwarf, i.e., it has a value of $v\,\sin{i}$ below our
detection limit. We estimate our detection limit to be $v\,\sin{i} =
3$\,km\,s$^{-1}$ as in \cite{RB07}. There we show that rotational
broadening of $v\,\sin{i} = 3$\,km\,s$^{-1}$ can marginally be
detected in spectra of this quality (the detection limit is set by the
spectral resolution of the data). We show a part of the FeH band in
2MASS1045$-$01 in the top panel of Fig.\,\ref{fig:spectra}. Although
many lines are strongly saturated, many sharp spectral features are
still visible.

With a slowly rotating L dwarf in hand, our fitting process is more
robust because it suffers much less from spectral mismatch.  We
measure $v\,\sin{i}$ in all other L dwarfs by fitting a scaled version
of the spectrum of 2MASS1045$-$01 to the target spectra. To control
instrumental effects, we took spectra of 2MASS1045$-$01 with both
instruments, one with HIRES and one with UVES. In all L dwarfs shown
in this paper, a good fit was achieved in the sense that the
difference between the target spectrum and the artificially broadened
and scaled template spectrum of 2MASS1045$-$01 was on the order of the
noise in the data. Examples of the fit quality are shown in the center
and bottom panels of Fig.\,\ref{fig:spectra}. The rich structure of
the FeH band also gives a good sense of the fit quality since
remarkable features appear in all spectra even at high rotation.

Projected rotation velocities $v\,\sin{i}$ are given in the fourth
column of Table\,\ref{tab:results}. We estimate our uncertainties to
be of the same order as our detection limit in slow rotators, i.e.,
$\Delta\,v\,\sin{i} = 3$\,km\,s$^{-1}$. This is probably too small in
cases of very rapid rotation when the spectral features become smeared
out. We checked the achieved $\chi^2$ fit quality in each object, from
which we estimate a conservative relative uncertainty of $\pm 10$\,\%
in the rapidly rotating L dwarfs. We do not provide the formal
uncertainties from the $\chi^2 \pm 1$ (or 4) values because in most
cases they lead to very low uncertainties (a few hundred m/s). This is
because the main source of mismatch between data and fit is of
a systematic nature rather than purely statistical. Our more conservative
estimate relative uncertainty of $10\,\%$ is always well beyond the
formal uncertainties. We note that even in the cases of strong
saturation our scaling method provides a very good match to the data
(see Fig.\,\ref{fig:spectra}).

In three cases, the fitting process yields uncertainties larger than
10\,\%. The data of 2MASS0913+18 (L3.0) and 2MASS0825+21 (L7.5) suffer
from low SNR. In the latter, fitting is particularly difficult because
the FeH band becomes strongly saturated in the late L-dwarfs, and dust
becomes an issue in such cool objects. This problem also arises in the
(high SNR) spectrum of 2MASS0255+47 (L8.0). We estimate that the
uncertainties in $v\,\sin{i}$ in these three objects are on a
20\,\%-level.

\begin{figure}
  \plotone{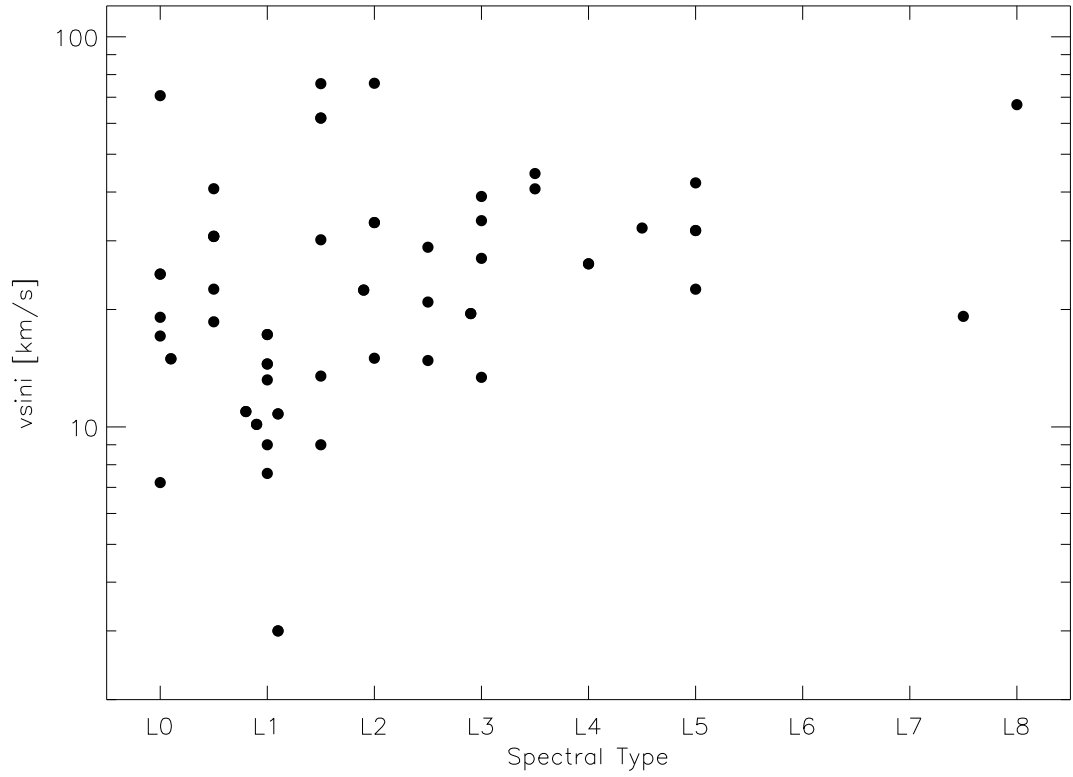}
  \caption{\label{fig:vsini}New measurements of $v\,\sin{i}$ as a
    function of spectral type.  $v\,\sin{i} = 3$\,km\,s$^{-1}$ is used
    as a lower limit.}
\end{figure}

Our data set consists of three parts, (a) HIRES observations, (b) new
UVES observations, and (c) UVES archive data.  From our sample of 45 L
dwarfs, 14 are contained in two of the three subsets, and one object
(2MASS0746$+$20) was observed in all three campaigns. In almost all
cases, the projected rotation velocities derived from different
spectra very accurately match each other, i.e. the difference is lower
than 2\,km\,s$^{-1}$. For the stars with more than one measurement, we
give the mean of the different measurements in
Table\,\ref{tab:results}.  This is also the value used in
Fig.\,\ref{fig:vsini}.

The only exceptions with differences
between different spectra appear for 2MASS1506$+$13 (L3, $v\,\sin{i} =
15$ and $24$\,km\,s$^{-1}$), and for 2MASS1507$-$16 (L5, $v\,\sin{i} =
28$ and $35$\,km\,s$^{-1}$). 

For 2MASS1506$+$13 we possess one spectrum taken with HIRES
(24\,km\,s$^{-1}$) and a second spectrum taken during our UVES
campaign (15\,km\,s$^{-1}$).  Both spectra have rather high SNR and we
feel confident that the result from each individual fit are within the
above mentioned uncertainties.  Thus, the spectra yield significantly
different results on rotational broadening on the order of
5\,km\,s$^{-1}$.  A detailed inspection of both spectra reveals that
indeed there is a visible difference between the two spectra taken
within only four weeks. We believe that this difference is due to the
excellent seeing conditions during the UVES observation of
2MASS1506$+$13 -- the one yielding the slower rotation velocity.
Seeing conditions were reported at a level below 0.6\arcsec during
that observation, while the slitwidth was 1\arcsec, which is on the
order of the seeing conditions we usually achieved. Assuming that for
the full observation the star was perfectly positioned in the slit,
the effective resolution of this exceptional observation would be
almost a factor of two higher than expected, leading to narrower
instrumental broadening and thus to a smaller rotational velocity. We
think that the exceptional quality of the seeing can fully explain the
difference compared to the result calculated from the HIRES spectrum.

In the second case of different $v\,\sin{i}$'s, 2MASS1507$-$16, we
possess two UVES spectra, one from our observation and one from the
archive. In this case, the two values ($28$ and $35$\,km\,s$^{-1}$)
are still marginally consistent and we see no obvious differences
between the spectra. Hence the difference is probably of purely
statistical nature (within the 1-sigma overlap of the error bars). 
Together with the 13 other objects in which the
results from different spectra (and different instruments) are within
only 2\,km\,s$^{-1}$, we feel very confident that instrumental effects
play no major role, and that our results are indeed within the
estimated uncertainties.

\subsubsection{Comparison to literature}
\label{sect:litcomp}

\cite{Bailer04} measured the projected rotation velocity in a sample
of L dwarfs by applying the cross correlation method (no measurements
of the H$\alpha$ line are given). We re-analyzed the data after
reduction of the raw data that we took from the ESO science archive,
applying the same strategy we used for our HIRES- and UVES-data.
H$\alpha$ measurements are given here for the first time. We first
performed a cross-correlation analysis following the method employed
by \citet[][we chose to use only the FeH band that was also contained
in their analysis]{Bailer04} in order to check differences due to our
new reduction of the data. We find that our results from the cross-
correlation analysis are fully consistent with the results given in
\citet{Bailer04}.

\begin{figure}
  \plotone{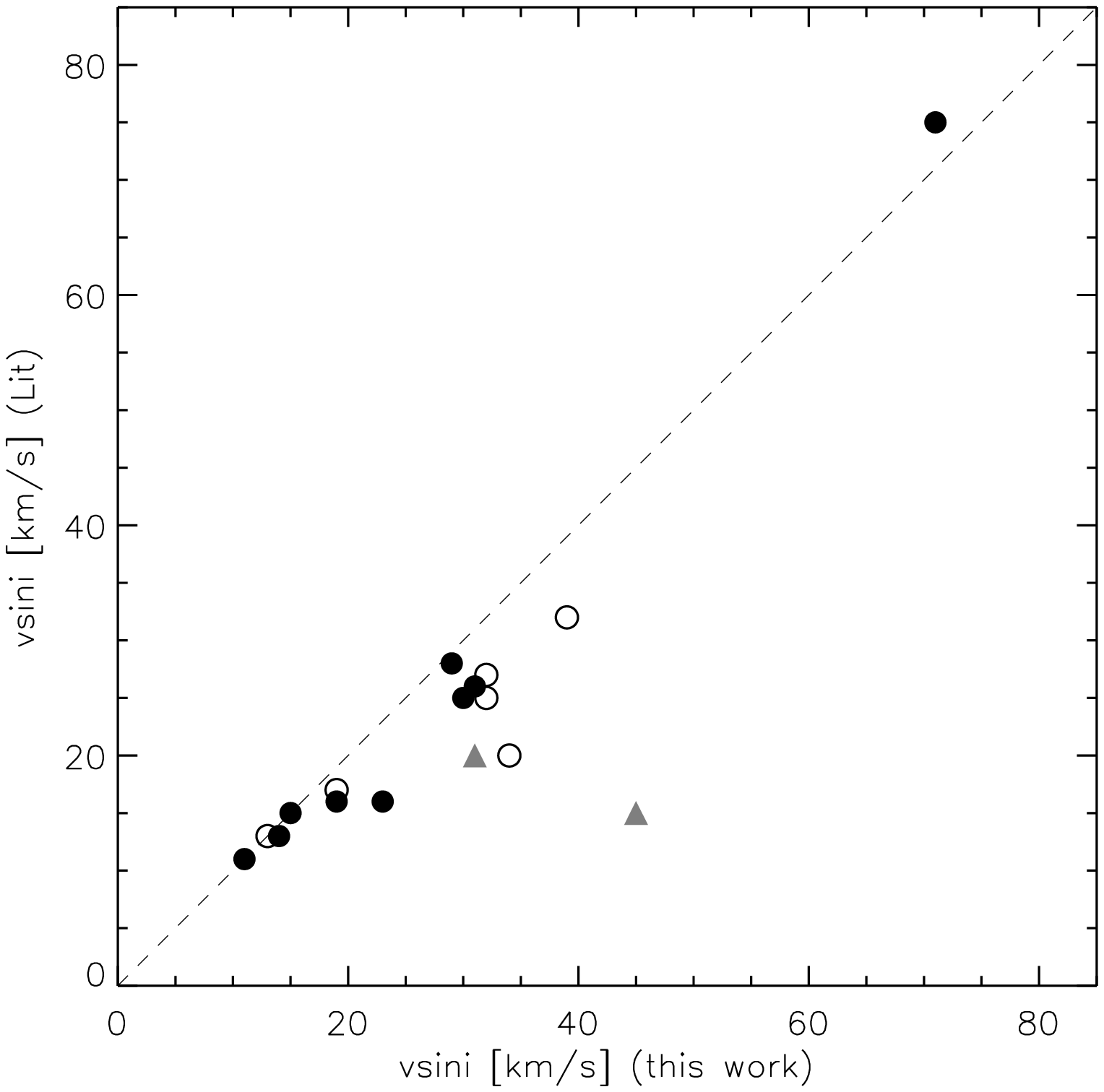}
  \caption{\label{fig:BJcomp} Comparison to \citet[][filled circles:
    spectral type earlier than L3; open circles: spectral type L3 and
    later]{Bailer04} and \citet[][grey triangles]{Schweitzer01}.  The
    dashed line marks the line of equality. }
\end{figure}

Next, we calculated the rotational velocities with our fitting
procedure. We compare the new results to the ones from the analysis of
\cite{Bailer04} in Fig.\,\ref{fig:BJcomp}.  The results are
comparable at slow rotation ($v\,\sin{i} \la 20$\,km\,s$^{-1}$) and
also in the most rapid rotator (80\,km\,s$^{-1}$). Between these
points, between 20-45\,km\,s$^{-1}$, there are several objects for 
which the cross-correlation yields smaller rotation velocities
than the direct fitting procedure does. We argue that our method
yields the more reliable results and we attribute the differences to
the two important improvements in our analysis: (a) we could use a
slowly rotating L1 object as a template, and (b) we used a well-known
spectral range with intrinsically narrow spectral lines in which we account
for the scaled depth of all absorption lines.

\citet{Mohanty03} measured activity and rotation in mid-M to L field
dwarfs, and \citet{Zapatero06} measure rotation in M, L and T dwarfs.
We have 4 objects in common with the sample of \citet{Mohanty03}:
2MASS1439+19, 2MASS1305-25 (Kelu-1), 2MASS0004-40 (LHS102B), and
2MASS0255-47 (DENIS0255-4700, L8).  Rotation velocities of the first
three objects are consistent within the uncertainties. However, our
result for the L8 object 2MASS0255-47 of $v\,\sin{i} =
67$\,km\,s$^{-1}$ is much larger than the result reported in
\citet{Mohanty03} of $v\,\sin{i} = 40$\,km\,s$^{-1}$. The sample of
\citet{Zapatero06} has three objects in common with our sample,
2MASS0036+18 (L3.5), and 2MASS2224-01 (L4.5), and also 2MASS0255-47.
Again, the results for the two mid-type L dwarfs are consistent.
However, \citet{Zapatero06} also report $v\,\sin{i} =
40$\,km\,s$^{-1}$ for 2MASS0255-47. This object is the coolest in our
sample, and our template spectrum may not provide as good a match to
the spectrum of 2MASS0255-47 as it does in earlier objects. In late-L
dwarfs, the depth of the FeH band is much weaker than in earlier L
dwarfs probably due to the presence of dust.  In this case, it is
difficult to disentagle the effects of saturation and rotational
broadening. Nevertheless, a careful inspection of the fit quality in
2MASS0255-47 shows that in our data a clear minimum in $\chi^2$
appears around $v\,\sin{i}$ of 60--70\,km\,s$^{-1}$, and that
$v\,\sin{i} = 40$\,km\,s$^{-1}$ provides a much worse fit in the FeH
band.  \citet{Mohanty03} and \citet{Zapatero06} used templates of
spectral types M5.5, M8, and T6.5 to determine $v\,\sin{i}$. We
believe that our analysis is probably more reliable because we tried
to match the line depths according to the target spectrum, and because
our template spectrum still is more similar to an L8 spectrum than
mid-M or late-T dwarf spectra are.

\citet{Schweitzer01} fitted model spectra to alkali lines of late-M
and early-L dwarfs measuring $T_{\rm eff}$, $\log{g}$, and
$v\,\sin{i}$. Two of their targets are also in our sample.
\citet{Schweitzer01} measured $v\,\sin{i} = 20$\,km\,s$^{-1}$ for
2MASS0746$+$20 for which we found $v\,\sin{i} = 31$\,km\,s$^{-1}$. For
2MASS0036$+$18 they report $v\,\sin{i} = 15$\,km\,s$^{-1}$, our value
is $v\,\sin{i} = 45$\,km\,s$^{-1}$. The two objects are also plotted
in Fig.\,\ref{fig:BJcomp}; a significant discrepancy exists between
our values and the ones measured by \citet{Schweitzer01}. We believe
that this is due to mismatches between the observed spectra and the
models used in their analysis. Fitting the heavily broadened alkali
lines in such cool objects is a problem, because pressure broadening
as well as the influence of dust is not fully understood \cite[see,
e.g.,][]{Johnas08, R07b}.

\section{Stars with Variable H$\alpha$ Emission}

\begin{figure*}
  \parbox{\textwidth}{
    \includegraphics[width=.19\hsize,clip=,bbllx=0,bblly=0,bburx=260,bbury=450]{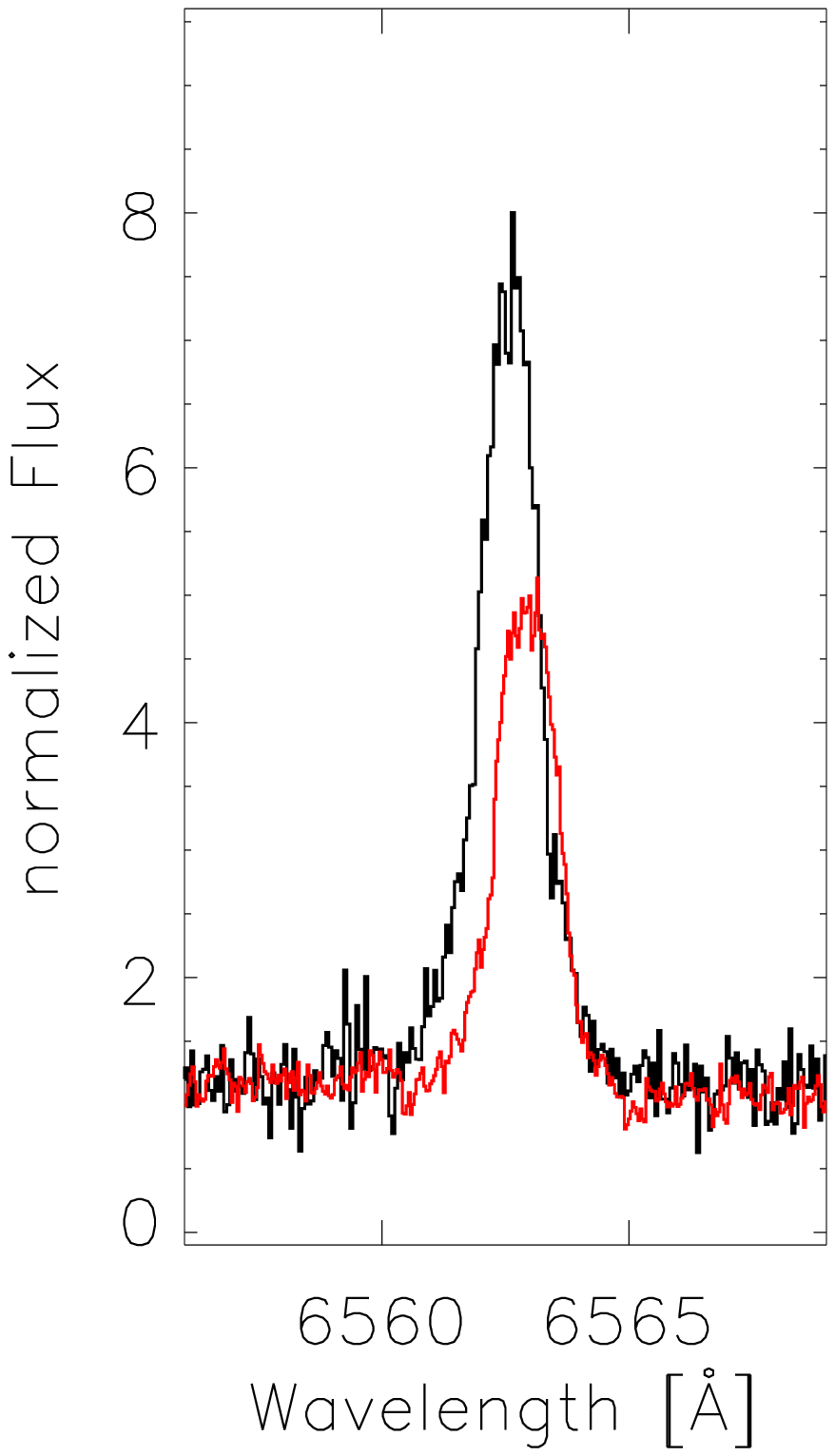}
    \includegraphics[width=.19\hsize,clip=,bbllx=0,bblly=0,bburx=260,bbury=450]{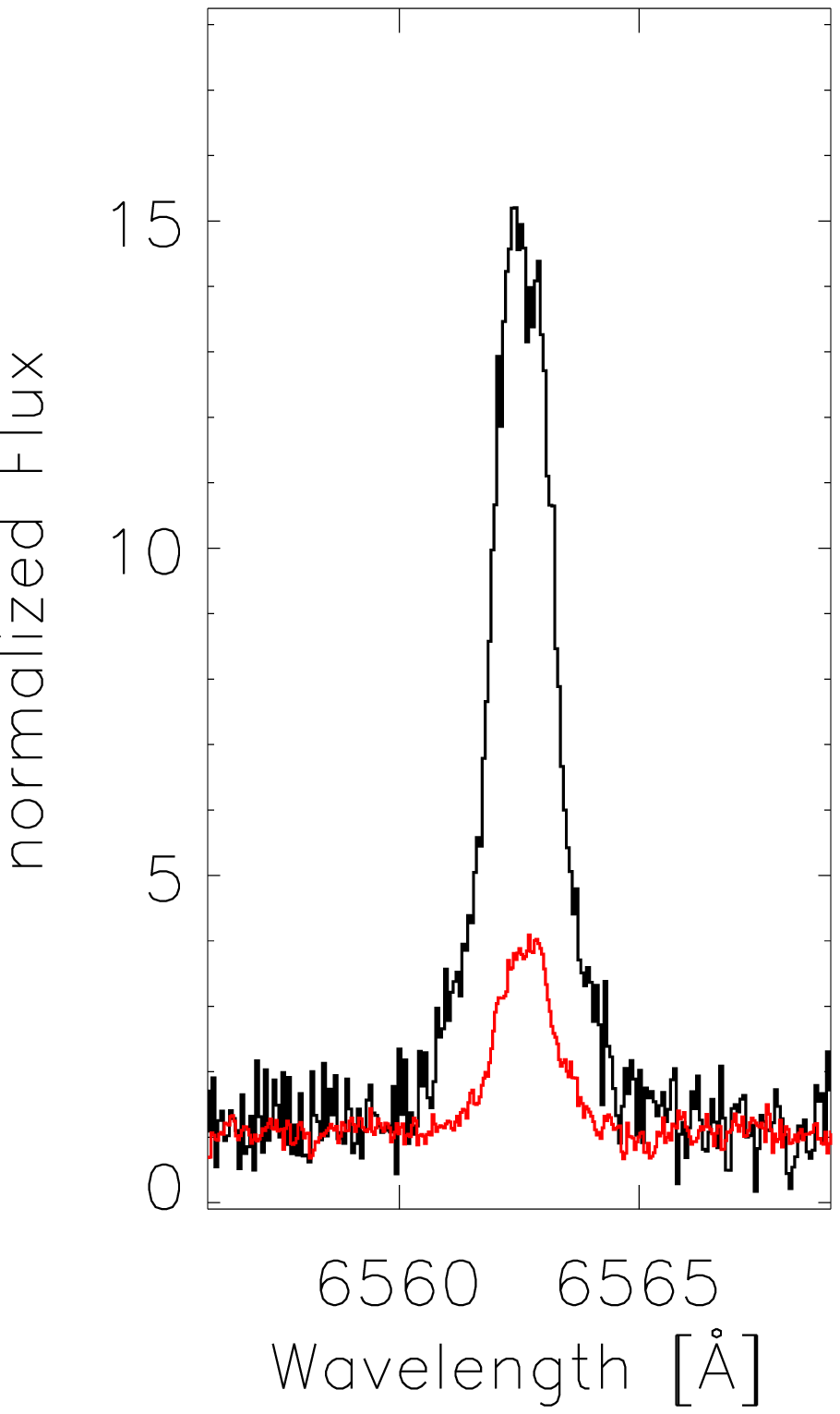}
    \includegraphics[width=.19\hsize,clip=,bbllx=0,bblly=0,bburx=260,bbury=450]{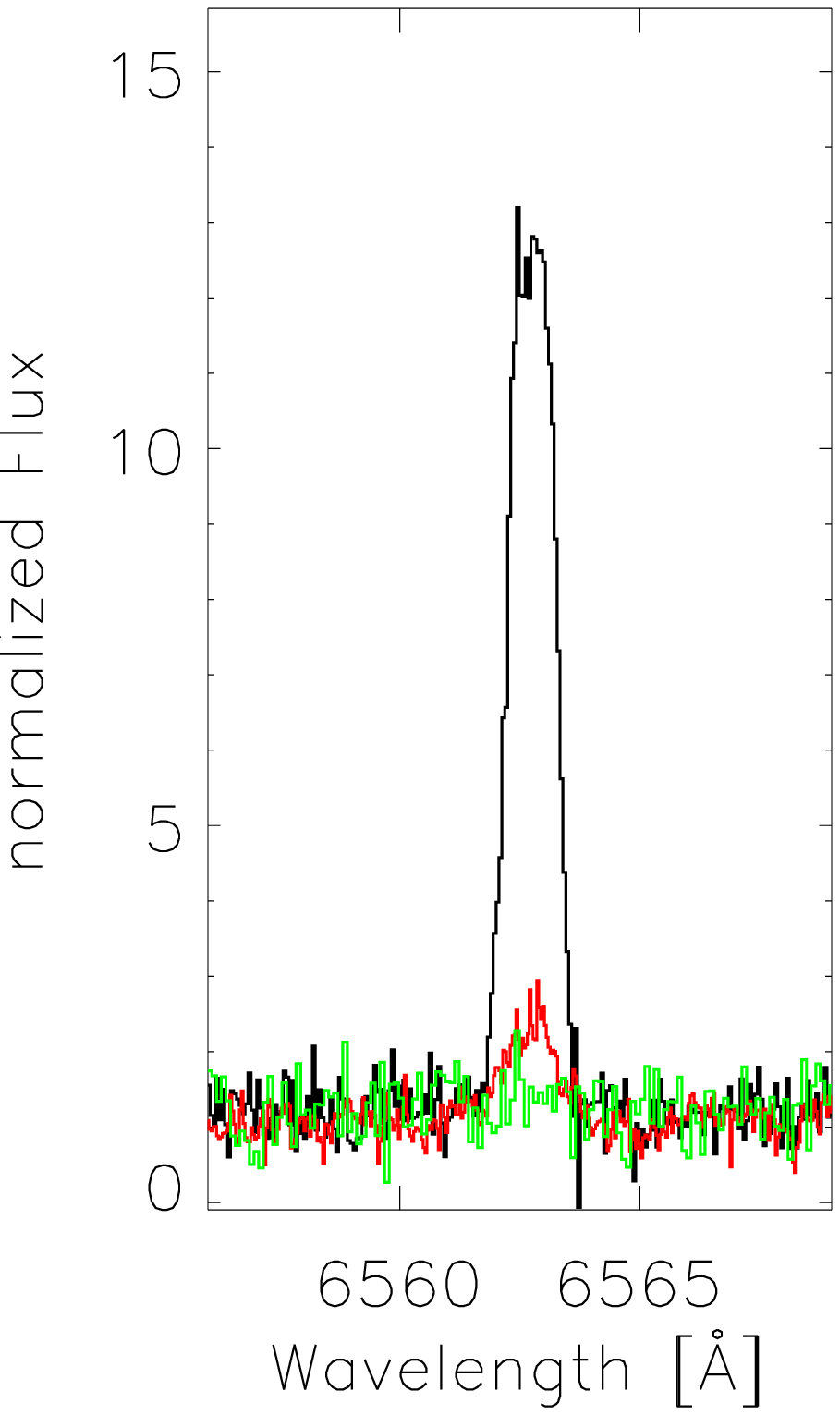}
    \includegraphics[width=.19\hsize,clip=,bbllx=0,bblly=0,bburx=260,bbury=450]{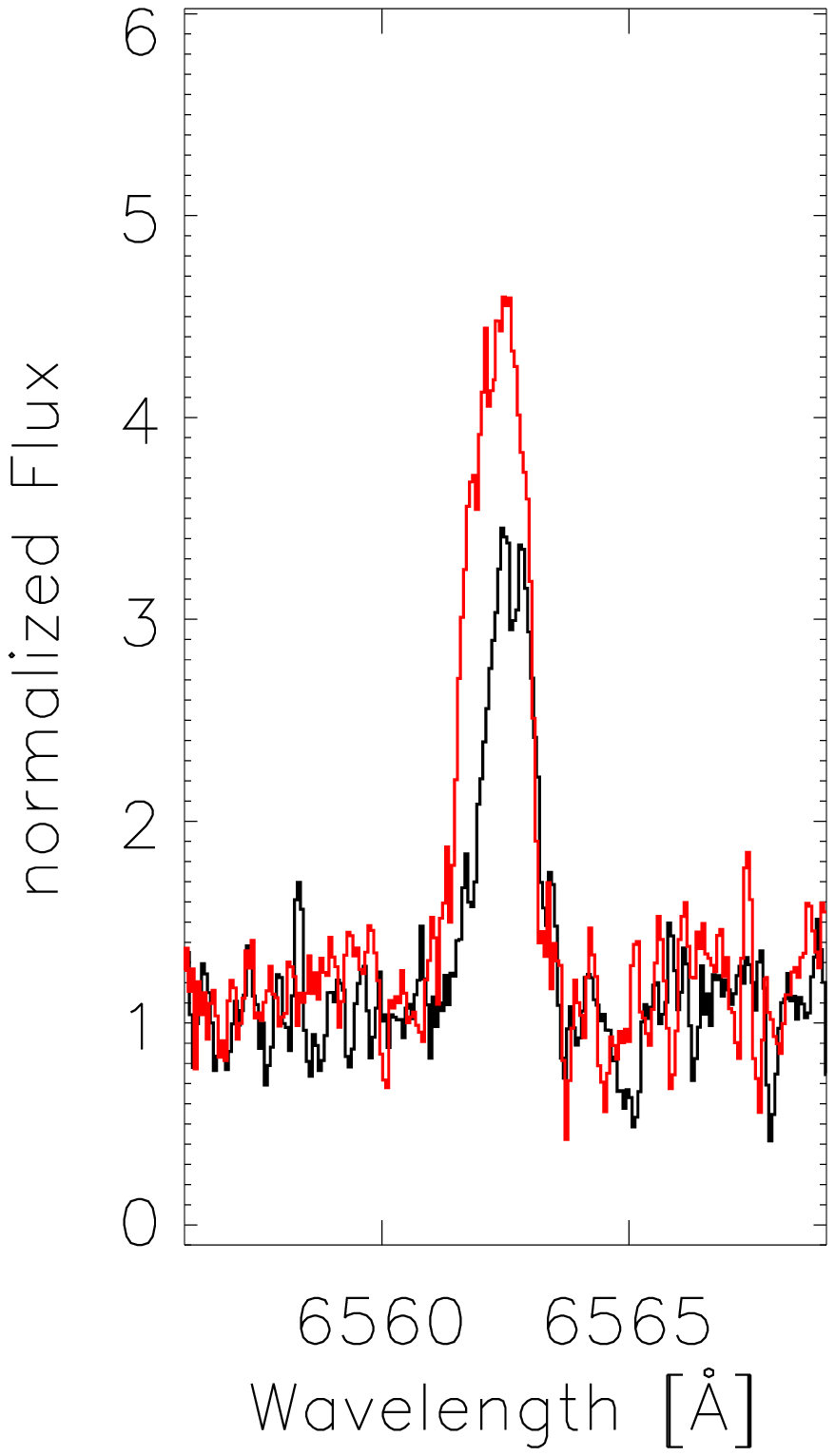}
    \includegraphics[width=.19\hsize,clip=,bbllx=0,bblly=0,bburx=260,bbury=450]{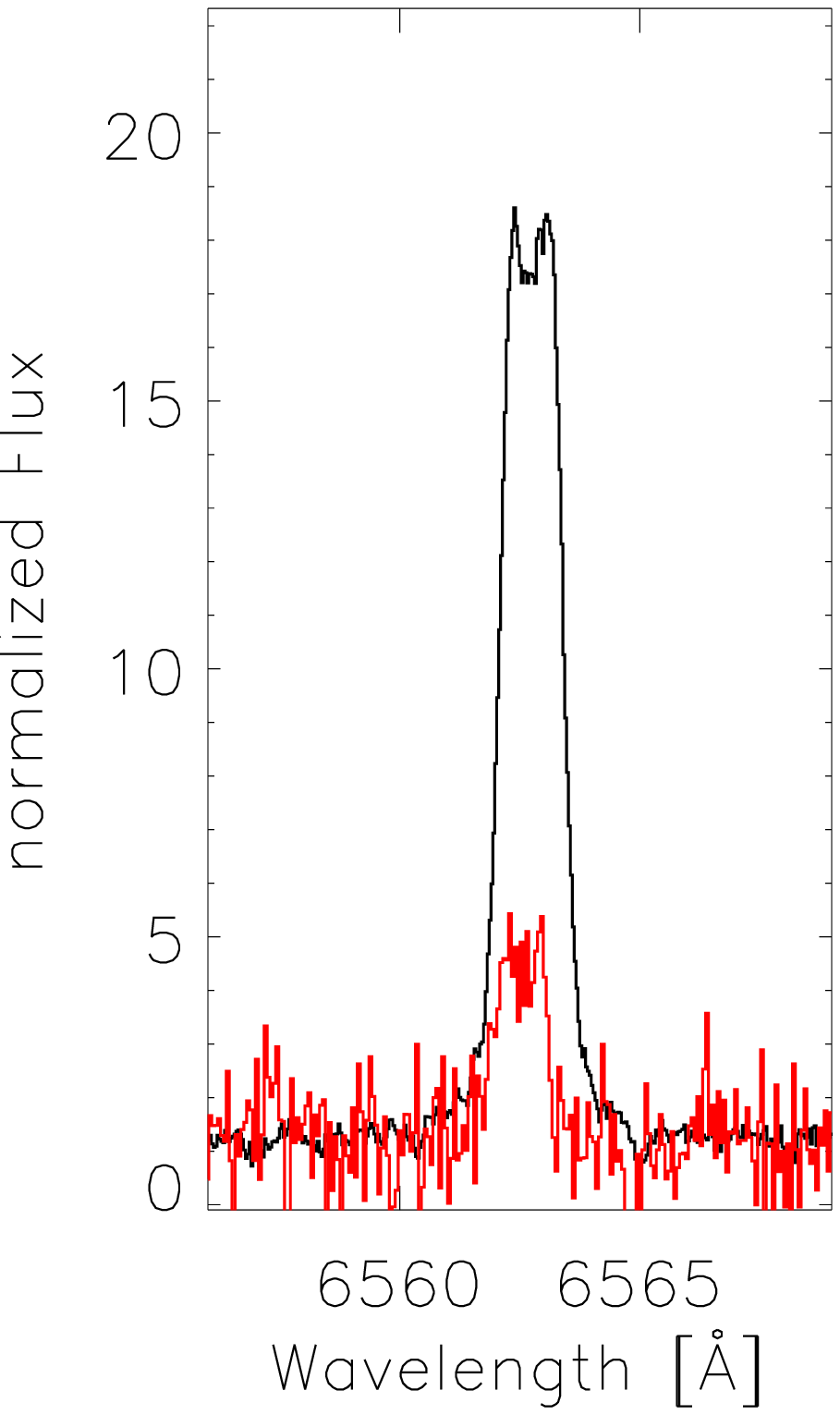}
  }
  \caption{\label{fig:flares}L-dwarfs with variable H$\alpha$
    emission. From left to right: 2MASS1731+27, 2MASS1221+02,
    2MASS1439+19, 2MASS1022+58, 2MASS1555-09. The spectrum of
    2MASS1022+58 was smoothed with a 3-pixel boxcar. Note that the
    scaling is different (continuum flux is normalized to unity).}
\end{figure*}

The sample of high resolution spectra we analyze in this paper is
taken from two different observatories obtained during several
observing runs. The goal was to obtain as many high quality L dwarf
spectra as possible, and to get multiple epochs in some of them.  We
can investigate time variability of objects for which we have multiple
observations; 15 objects were observed more than once, one object was
observed three times. In Table\,\ref{tab:halpha}, we list observations
of the 8 targets that show significant variability in H$\alpha$.  This
observed fraction implies that in this very heterogeneous sample the
variability of L dwarfs is $53\pm13\,\%$.  Fig.\,\ref{fig:flares}
displays the most obvious cases of variable H$\alpha$ emission,
probably related to flaring.

\section{Results}
\label{sect:results}

\subsection{Activity}

\begin{figure}
  \plottwo{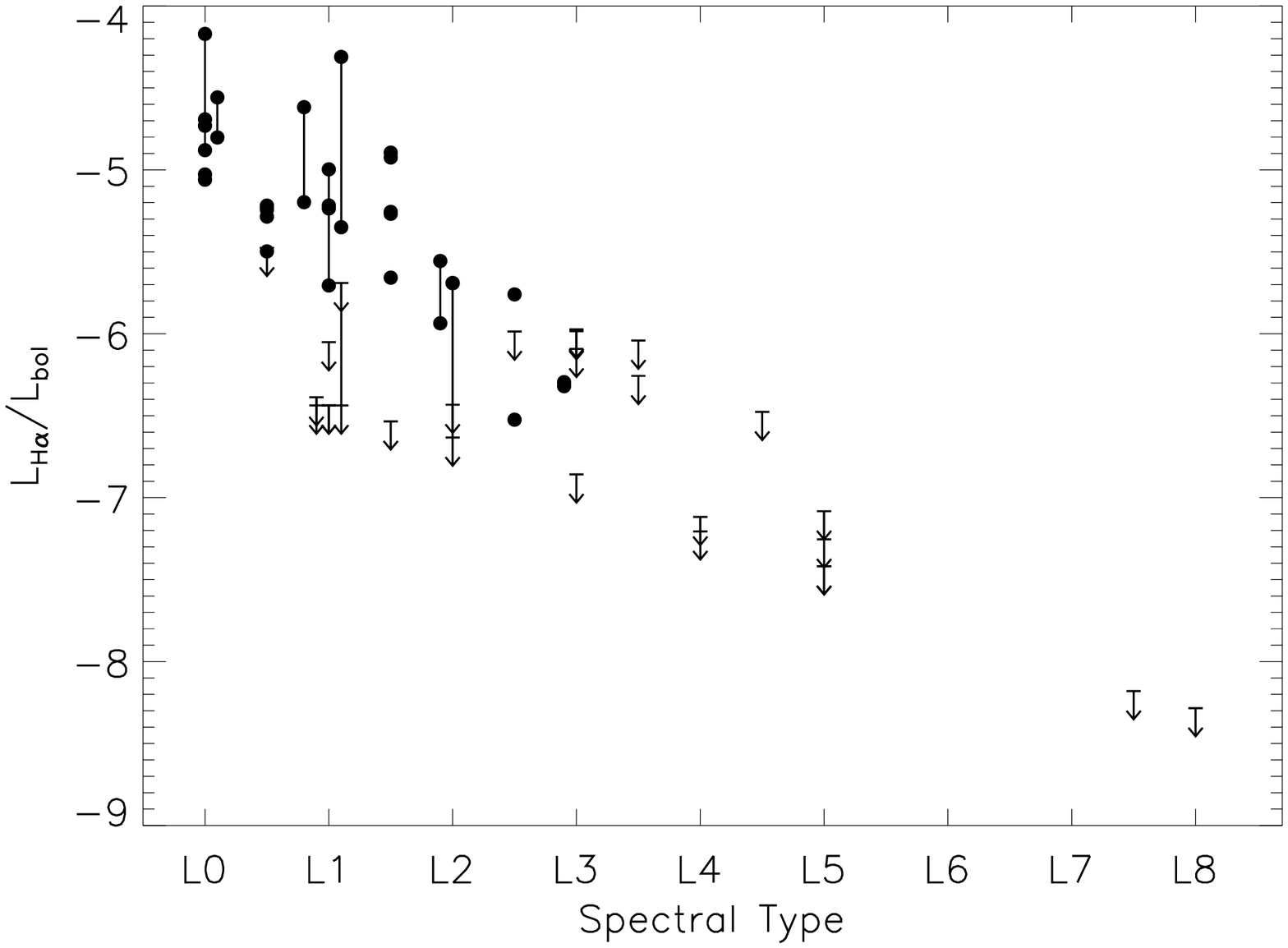}{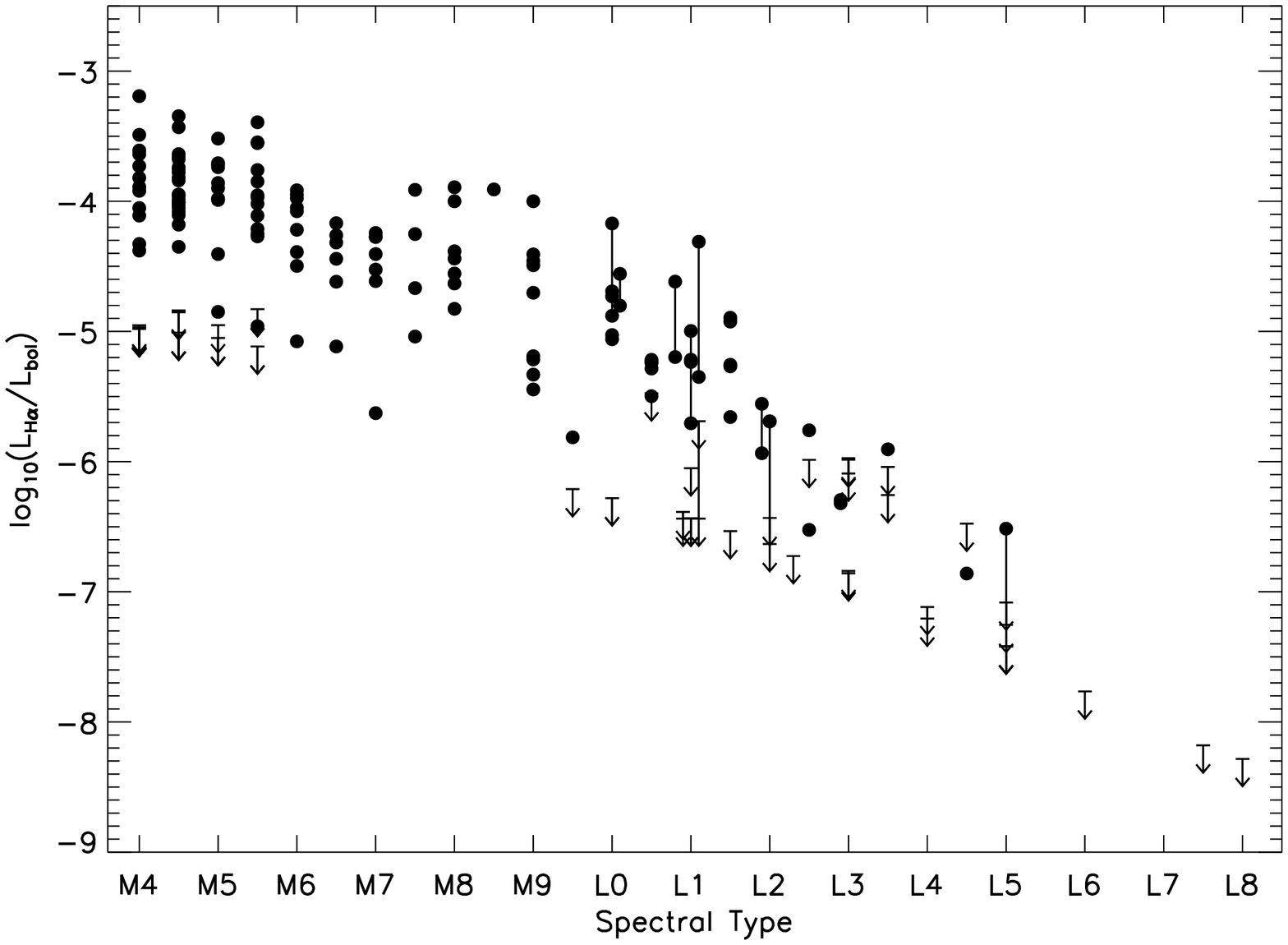}
  \caption{\label{fig:Halpha}Measurements of chromospheric activity in
    terms of $L_{\rm{H}_\alpha}/L_{\rm{bol}}$. Different observations
    of the same object are connected with solid lines. Upper limits
    are indicated with downward arrows. Left: Measurements from our
    new sample; right: new measurements together with literature
    values.}
\end{figure}

We plot the normalized H$\alpha$ luminosities as a function of
spectral types in Fig.\,\ref{fig:Halpha}. Results from our new
observations are shown in the left panel. In the right panel, we show
them together with the data from \citet{Delfosse98}, \citet{RB07}, and
\citet{Mohanty03}.  We have four objects in common with the sample of
\citet{Mohanty03}, in none of the four did we detect H$\alpha$
emission, while \citet{Mohanty03} did detect H$\alpha$ in
2MASS0004$-$40 (LHS~102B).

H$\alpha$ emission appeared in 31 of the 61 spectra (51\,\%) that we
took in 45 stars. 23 of the 45 ($51\pm8\,\%$) objects show H$\alpha$
in emission in at least one spectrum. For one of them we also possess
a spectrum in which H$\alpha$ emission was below the detection level.
In agreement with \cite{Mohanty03}, \cite{West04}, and
\cite{Schmidt07}, we find that the normalized H$\alpha$ luminosity is
decreasing with later spectral type. This effect is likely due to the
growing neutrality of cool dwarfs' atmospheres \citep{MBS02}.
Nonetheless, activity and flaring events were also found in mid-type
L-dwarfs: \citet{Mohanty03} observed H$\alpha$ emission in high
resolution spectra of objects as late as L5 (contained in this
analysis and included in the right panel of Fig.\,\ref{fig:Halpha}).
\citet{Hall02} and \citet{Liebert03} found H$\alpha$ emission in the
L5 dwarfs 2MASS~J1315309$-$264951 and 2MASS~J01443536$-$071642,
respectively.  Other tracers also indicate that L dwarfs are capable
of generating activity. For example, radio emission was detected in
two L dwarfs, 2MASS0523$-$14 (L2.5) and 2MASS0036$+$18
\citep[L3.5,][]{Berger06}.

\begin{figure}
  \plotone{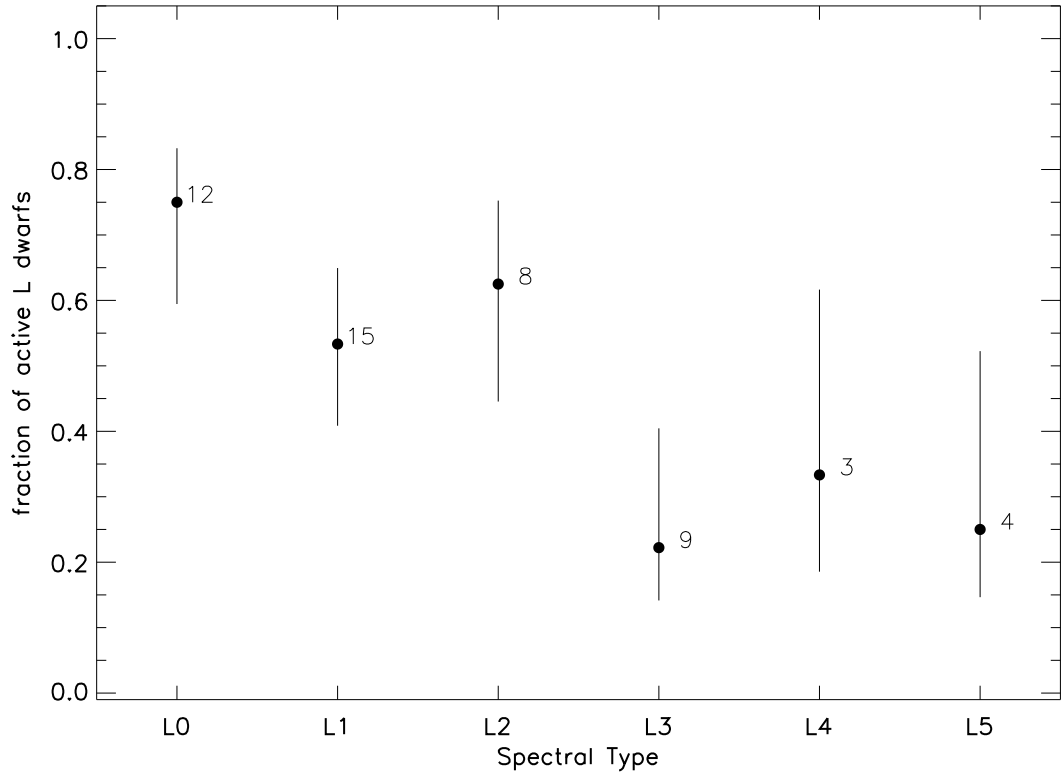}
  \caption{\label{fig:Lfraction} Fraction of active L-dwarfs per
    spectral bin. The total number of objects per bin is indicated.}
\end{figure}

In Fig.\,\ref{fig:Lfraction}, we show the fraction of active objects
per spectral type bin from L0 to L5. Our high resolution L dwarf data
is combined with the sample of \citet{Mohanty03}, who used data of
comparable quality around the H$\alpha$ line. The total number of
objects per spectral bin is indicated next to the points in the plot.
Fig.\,\ref{fig:Lfraction} confirms that the (measured) fraction of
active objects decreases towards cooler temperatures. This plot can be
compared to Fig.\,1 in \citet{West04} and to Fig.\,5 in
\citet{Schmidt07}. In both papers low resolution data was used, and
the detection limit of H$\alpha$ emission is lower than in our data.
Our results show fractions that are 10--20\,\% higher than the ones
from \citet{West04} and \cite{Schmidt07} in all spectral bins. We
conclude that low levels of H$\alpha$ activity are probably generated
by the majority of L-dwarfs, and with higher detection sensitivity the
observed fraction of active stars becomes higher.

\citet{Mohanty03} find a very sharp difference in activity between M
stars up to spectral type M9 and cooler objects, i.e., normalized
H$\alpha$ luminosities drop about two orders of magnitude around a
spectral type of M9 \citep[see also][]{Liebert03}. Our larger sample
shows that this was a selectional bias and not a physical effect.  We
clearly see that $L_{\rm H{\alpha}} / L_{\rm{bol}}$ gradually
decreases from mid-M to mid-L spectral types, and that no sudden break
occurs.

\subsection{Activity and Age}

In Sun-like stars, activity is strongest when stars are young and
rotating most rapidly. During their evolution, rotation is braked and
the level of activity decreases over time. Do the same mechanisms also
work in mid-M and L-dwarfs, which are certainly fully convective? Much
less is known about the mechanism generating the presumably required
magnetic fields or the way they are converted into atmospheric heating
via dissipative mechanisms, and then to a braking torque by a magnetic
wind.  In order to investigate the dependence of H$\alpha$ emission on
age, we employ the measurements of tangential velocity given in
\citet{Schmidt07}. For objects of our sample that are contained in
that work, we give the values of $v_{\rm{tan}}$ in the last column of
Table\,\ref{tab:results}.

Although tangential velocity does not capture the full space motion,
it is a convenient measure which provides a preliminary picture for
the statistical investigation of a large sample. In general, young
stars have very low total space velocity \citep[e.g.,][]{Chiba00} and
any detection of motion on the order of, say, 30\,km\,s$^{-1}$
indicates that the object is probably not a member of the young
population. For a deeper investigation of the velocity distribution,
we must measure radial velocities from our data and calculate space
motions, which we plan to do in a separate paper.

\begin{figure}
  \plottwo{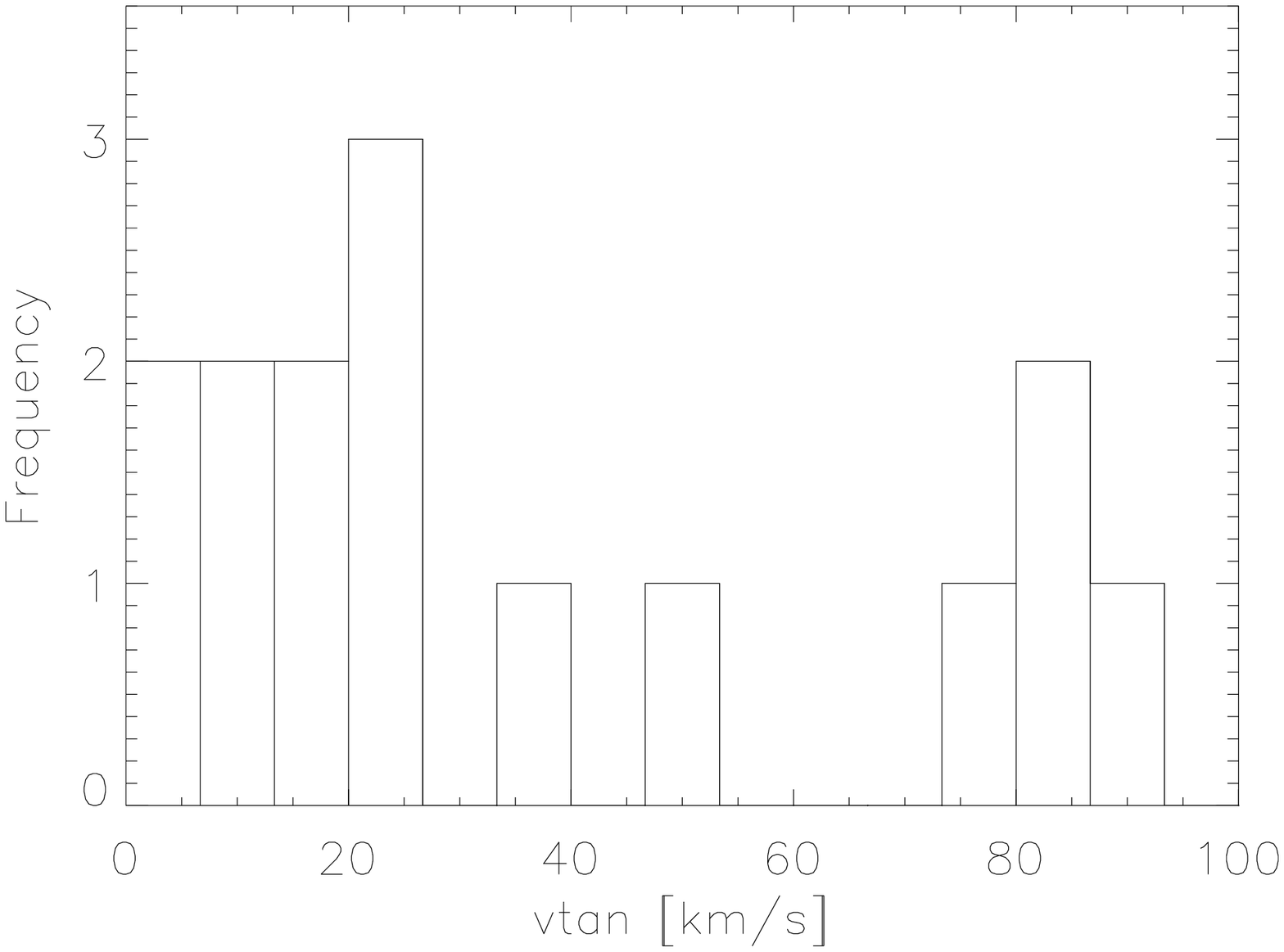}{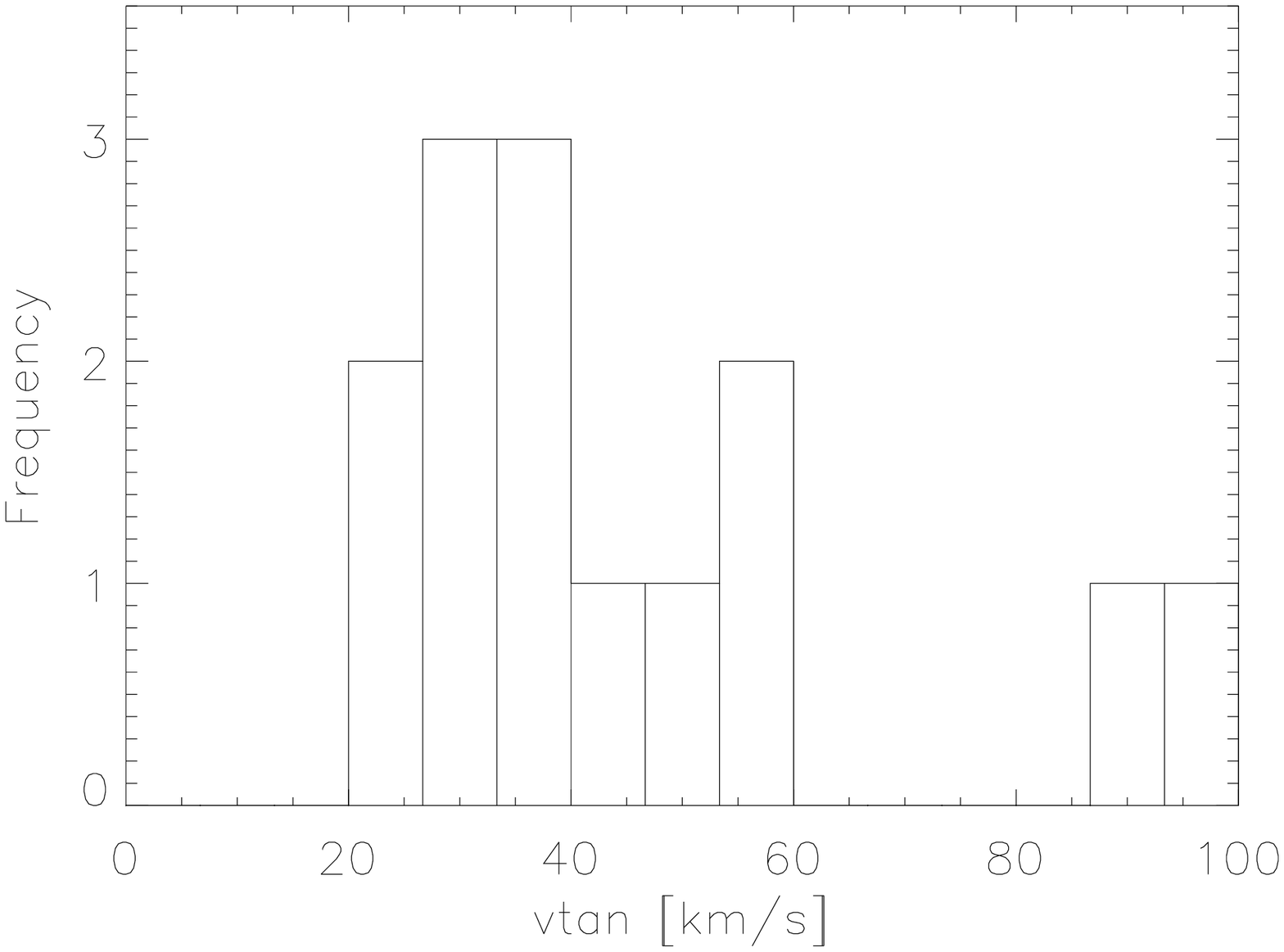}
  \caption{\label{fig:histo_vtan}Histograms of active (left panel) and
    inactive L dwarfs (right panel) in tangential velocity.}
\end{figure}

We plot the distribution of tangential velocities in active and
inactive (no H$\alpha$ detection) objects in the left and right panels
of Fig.\,\ref{fig:histo_vtan}, respectively. Both distributions cover
the entire range between $v_{\rm{tan}} = 0$ and 100\,km\,s$^{-1}$. The
most obvious difference is that 6 objects with detected H$\alpha$
emission have values of $v_{\rm{tan}} < 20$\,km\,\,s$^{-1}$, while
none of the ``inactive'' objects has such a low tangential velocity.
The majority of the active objects are in the lower half of observed
tangential velocities as well.  We performed a $\chi^2$-test on the
two distributions asking whether both distributions could be due to
the same underlying population.  We find that the hypothesis of
identical underlying distributions can be rejected only at a
1$\sigma$-level (the probability that two observations of the same
distribution yield larger differences than our is 19\,\%). Thus,
although the lack of ``young'' inactive stars is suggestive, it is not
statistically significant in our sample.

Another way of investigating stellar activity is by measuring radio
emission, which is believed to be generated through gyrosynchrotron
emission in the presence of strong magnetic fields. Radio observations
in L dwarfs were carried out by \cite{Berger06}, who found radio
emission in the two L-dwarfs 2MASS0523$-$14 (L2.5) and 2MASS0036$+$18
(L3.5). Both objects are contained in our sample: 2MASS0523$-$14 is
kinematically young and shows H$\alpha$ emission, but we do not detect
H$\alpha$ emission in the kinematically old object 2MASS0036$+$18. The
detection of radio and H$\alpha$ emission in 2MASS0523$-$14 supports
the idea that both are generated via the same mechanism but
2MASS0036$+$18 does not. \citet{Berger05} investigated this object in
a multi-wavelength campaign; they detected neither H$\alpha$ nor X-ray
emission. They concluded that the atmospheric conditions in L dwarfs
are markedly different from early-type stars and even M dwarfs, which
show a well-defined X-ray/radio correlation. \citet{Berger05} also
found a periodicity in the radio emission of 2MASS0036$+$18 of 3\,h,
which would mean a rotation velocity of $\approx 37$\,km\,s$^{-1}$ --
we find $v\,\sin{i} = 45$\,km\,s$^{-1}$. Given the large uncertainties
in the radius determination, both values are still consistent with
each other, but indicate that if the radio period is indeed due to
rotation, the object is seen at high inclination.

\subsection{Rotation}

The projected rotation velocities $v\,\sin{i}$ of our sample are shown
as a function of spectral type in Fig.\,\ref{fig:vsini}. For 27 L
dwarfs, $v\,\sin{i}$ is reported here for the first time. In
Fig.\,\ref{fig:vsiniall} we show our data together with data available
from the literature \citep{Delfosse98, Mohanty03, RB07}. We augment
the plot by results from \citet{RB06a} who investigated two presumably
very old late M- and L-type subdwarfs, LSR\,1610$-$0040 (d/sdM7) and
2MASS0532$+$8246 \citep[sdL7, spectral types from][]{Burgasser07}. No
significant rotation was found in the M7 object LSR\,1610$-$0040, but
the L7 subdwarf 2MASS0532$+$8246 still shows very rapid rotation of
$v\,\sin{i} \sim 65$\,km\,s$^{-1}$. We plot these objects as squares
(including uncertainties) in Fig.\,\ref{fig:vsiniall}. Both results
are consistent with the lower envelope of minimum rotation visible in
that Figure, providing a strong argument for the lack (or weakening)
of rotational braking in cooler objects, even if they are very old. At
a minimum, therefore, spindown times become very long for very cool
objects. We also add the three components of the triple system
LHS~1070 \citep{RS07}. The system consists of two ultra-cool M8/M9
objects rotating at identical rotation speed and a more massive M5.5
component that rotates significantly more slowly.  These three objects
are plotted as stars.

We plot the ``young'' sample as blue symbols in
Fig.\,\ref{fig:vsiniall}, the ``old'' sample is shown with red
symbols. Objects for which no age information is available are shown
as open symbols. Clearly, the upper left quadrant of
Fig.\,\ref{fig:vsiniall} is occupied by the ``young'' sample while
near the envelope of minimum rotation velocity objects of the ``old''
sample predominate. The two presumably very old sub-dwarfs and the
three components of LHS~1070 also follow this trend. There is no
indication that our sample lacks old objects, and it shows that the
lack of slowly rotating L dwarfs is real.

The lower envelope of (projected) rotation velocities describes
minimum rotation close to zero at spectral type M5, $\sim
7$\,km\,s$^{-1}$ at L0, and around 20\,km\,s$^{-1}$ at spectral type
L5. \citet{Zapatero06} measured rotation velocities in T dwarfs at
lower spectral resolution. Their data show the same trend of higher
minimum rotation velocities at later spectral type. Among the
T~dwarfs, \citet{Zapatero06} found projected rotation velocities
$v\,\sin{i}$ around 30\,km\,s$^{-1}$. This result suggests that the
envelope of lower limits to rotation may flatten in the T dwarfs, and
the lack of more rapid rotators would be a change if confirmed in a 
larger sample. 
We are quite skeptical about a comparison of rotational 
velocities from very 
different spectral resolutions, however, and would
not put much stock in either conclusion for now.

It is important to realize that although we can measure only the
\emph{projected} rotation velocity, real rotation velocities can only
be higher -- not lower. Hence the effect of detecting only rapid
rotators among the L dwarfs is certainly not due to a projection
effect. The only exceptionally slow object among the rapidly rotating
L dwarfs is 2MASS1045$-$01. In that object, we did not detect any
rotational broadening.  We cannot say whether this is due to a
projection effect or if 2MASS1045$-$01 is indeed a slow rotator. To
find only one object at very low inclination, however, is
fully consistent with the assumption that \emph{all} L dwarfs in our
sample are rapid rotators.

\subsubsection{Rotational evolution of L dwarfs}

L dwarfs significantly cool during their evolution, so that L dwarf
evolutionary tracks in a rotation/temperature (or
$v\,\sin{i}$/spectral type) diagram do not translate into vertical
lines (as for hotter stars with essentially constant temperature).
Brown dwarfs gravitationally contract during their whole lifetime,
and hence continuously diminish their moment of inertia, leading to
rotational acceleration (although most of this occurs in the first Gyr).

\begin{figure*}
  \mbox{
    \includegraphics[width=.69\hsize,bbllx=0,bblly=0,bburx=630,bbury=335]{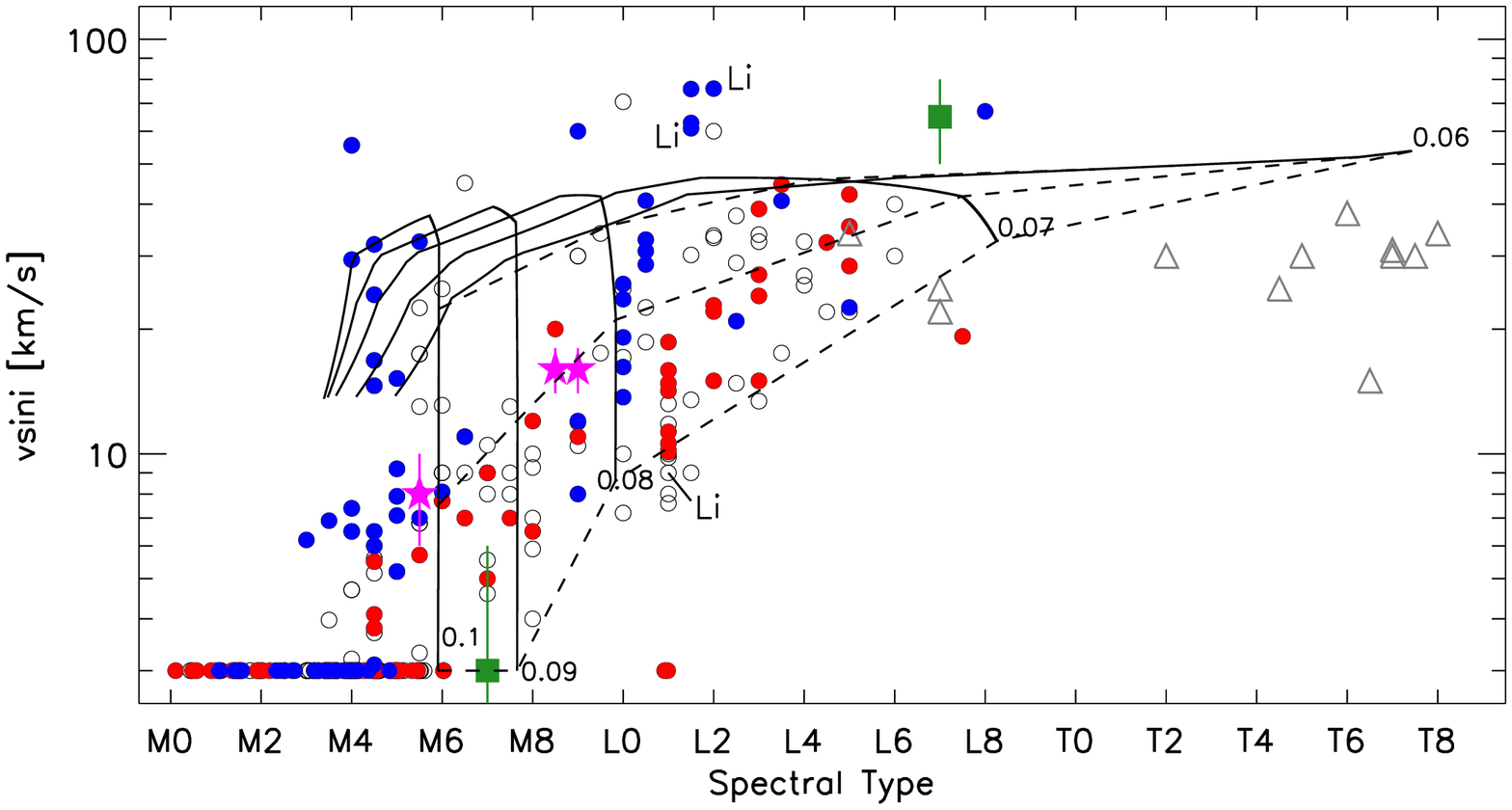}
    \includegraphics[width=.28\hsize,bbllx=100,bblly=-20,bburx=460,bbury=450]{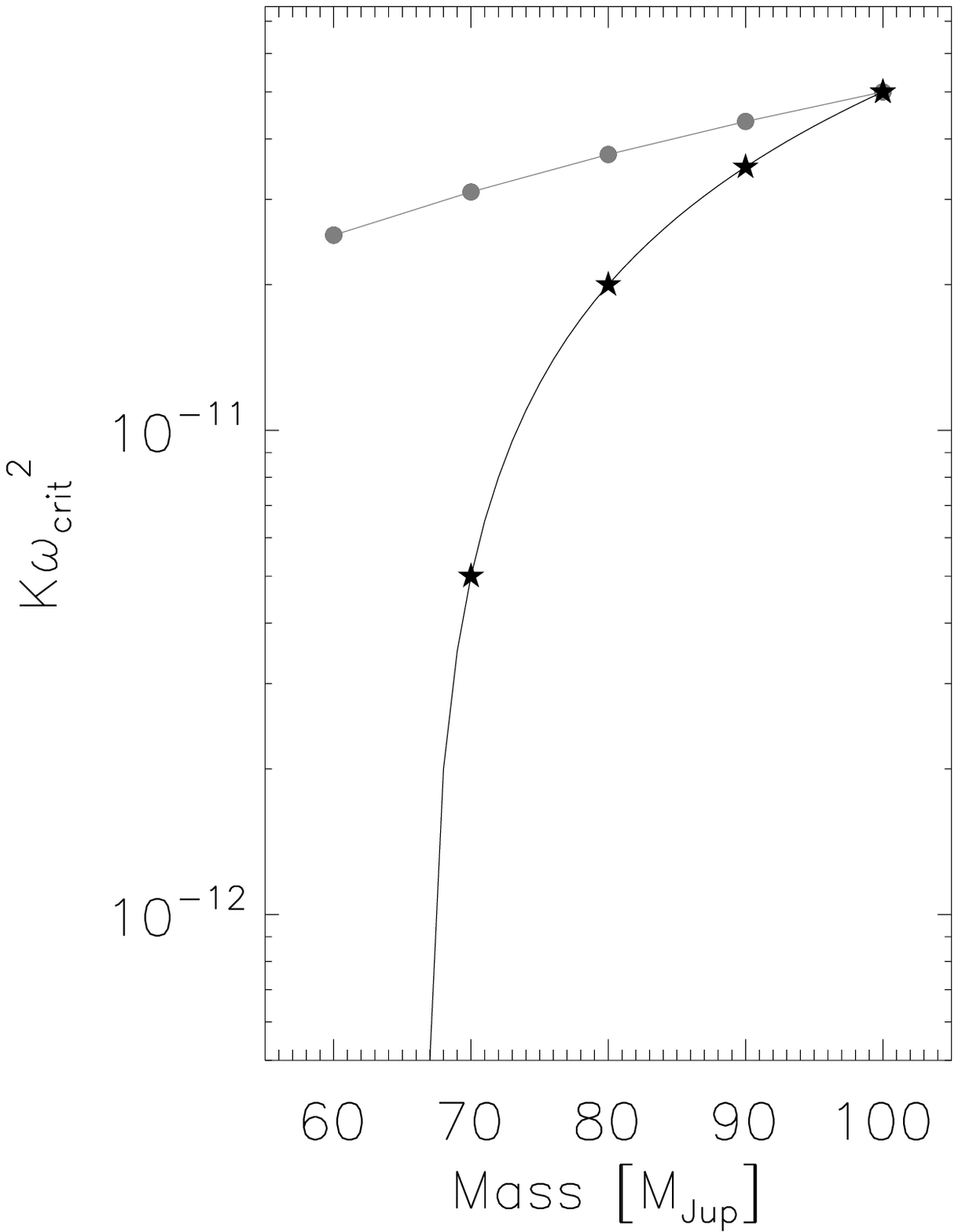}
  }
  \caption{\label{fig:vsiniall}\emph{Left:} Our measurements of
    $v\,\sin{i}$ together with results from the literature.
    $v\,\sin{i} = 3$\,km\,s$^{-1}$ is used as a lower limit. Circles
    are from this work, \citet{Delfosse98}, \citet{Mohanty03}, or
    \citet{RB07}.  Filled blue circles are probably young, filled red
    circles old.  Open circles are used for objects with no age
    information. Magenta stars indicate the three members of LHS~1070
    \citep{RS07}, filled green squares show the two subdwarfs
    2MASS~0532$+$8246 and LSR~1610$-$0040 \citep{RB06a}, and open grey
    triangles are used for measurements from \citet{Zapatero06}.
    Objects with Li detections are marked with 'Li'. The solid lines
    mark evolutionary tracks for objects of 0.1, 0.09, 0.08, and
    0.07\,M$_{\sun}$ (see text), dashed lines mark ages of 2, 5, and
    10\,Gyrs (from upper left to lower right).  \emph{Right:} Scaling
    of the magnetic wind-braking law.  Grey dots show the scaling
    according to $\tau \propto M^{-2/3}$, the black line shows our
    scaling. Stars mark the four points used for the lines in the left
    panel (see text).}
\end{figure*}

\citet{Zapatero06} have calculated the effects of acceleration by
contraction and they find a picture that is qualitatively consistent
with their rotation velocities. They do not allow for any wind
braking; their calculations start at a rotation velocity of
9\,km\,s$^{-1}$ at an age of 1\,Myr. In this scenario, low-mass
objects are starting as slow rotators. They gain angular momentum
through contraction and accelerate as they age. Thus, only very few
(extremely young) \emph{slow} rotators would be observed while many
more rapid rotators could be expected among the (old) field objects.
Slowly rotating L dwarfs would be young but of very low mass, which
could explain the rising slope of minimum rotation velocity among the
field objects.  However, one feature of this scenario is that all old
objects are more rapidly rotating! We can exclude this prediction from
the fact that we observe rapid rotators to be preferentially young,
and that the lower envelope of the slowest rotators per spectral type
is occupied by old objects (Fig.\,\ref{fig:vsiniall}).

If we assume that rotational braking in L dwarfs is similar to wind
braking in more massive stars, the wind braking law in low-mass
objects appears to be less effective than in early-M stars. The fact
that in Fig.\,\ref{fig:vsiniall} rotational evolution apparently goes
from young rapid rotators to old slow rotators strongly suggests net
rotational braking instead of net acceleration. In the following, we
search for a scenario to explain the main ingredients of
Fig.\,\ref{fig:vsiniall}. The main features are a) the shape of the
lower envelope of minimum rotational velocities with higher minimum
rotation at later spectral type, b) the distribution of rotation
velocities among the young and the old populations, c) the rotation
velocities of the three (coeval) members of LHS~1070, and d) the two
presumably old sub-dwarfs. To explain the net rotational braking, we
need to introduce a braking law, and the rising slope of minimum
rotational velocities suggests that this braking law is
mass-dependent.

For the form of the braking law, we follow the models of
\citet{Chaboyer95} and \citet{Sills00}. The angular momentum loss is
assumed to be proportional to some power of the angular velocity
$\omega$ \citep{Mestel84, Kawaler88}. This power law depends on the
magnetic field geometry. \citet{Chaboyer95} and \citet{Sills00} give a
prescription for the wind-braking law if the angular velocity $\omega$
is higher than a critical angular velocity $\omega_{\rm crit}$
\citep[this was required mainly to explain angular momentum loss in
the pre-main sequence, see][and references therein]{Chaboyer95}.

\begin{equation}
  \label{eq:braking}
  \frac{dJ}{dt} = -K \omega_{\rm crit}^2 \omega \left(\frac{R}{R_\sun}\right)^{0.5} \left(\frac{M}{M_\sun}\right)^{-0.5}
\end{equation}

The critical velocity is thought to be the value where the rotation
period becomes much smaller than the convective overturn time, i.e.,
$Ro = P/\tau_{\rm conv} \la 0.1$. In early-M stars, this happens at
relatively low rotation velocities on the order of a few km\,s$^{-1}$.
In late-M and L dwarfs, critical velocities are even lower because
$\tau_{\rm conv}$ quickly grows larger at lower masses. Hence our
objects are probably all rotating faster than critical.
\citet{Sills00} use a scaling of $\omega_{\rm crit}$ with mass as
introduced by \citet{Krishnamurti97}:

\begin{equation}
  \label{eq:omegacrit}
  \omega_{\rm crit} = \omega_{\rm crit \sun} \frac{\tau_\sun}{\tau}
\end{equation}

with $\tau$ the convective overturn timescale. The mass dependence of
angular momentum loss follows from observations in open clusters and
could possibly be explained by mass-dependent magnetic topologies due
to different magnetic field generation mechanisms. As a first
approximation for the convective overturn timescale $\tau$, we assume
$\tau \propto M^{-2/3}$ \citep[cp][]{RB07}. We calculate rotational
braking in low-mass objects taking into account contraction and a
wind-braking law as introduced above. We take the radii from
evolutionary tracks of \citet{Baraffe98} and \citet{Chabrier00},
conversion from temperature to spectral type was done using the
calibration of \citet{Golimowski04}. For our model, we assume that
rotational evolution starts at an age of 5\,Myr \citep[about the age
where disk-locking ends;][]{Strom89}. All our test objects start with
rotation velocities of 10\,km\,s$^{-1}$, which is motivated by the
projected rotation velocities in low-mass stars and brown dwarfs
observed during their T~Tauri phase (\citet{Mohanty05}, see also
\citet{Basri02}). Calculating the rotational evolution for this simple
case we indeed find more rapidly rotating stars at later spectral
classes; at a given age, very low mass objects are not as strongly
braked as the heavier objects.  However, such braking would still be
strong enough to decelerate all objects to essentially zero rotation
after 10\,Gyrs, which clearly is not observed.

In order to reproduce the observed situation, we adjust the scaling of
$K \omega_{\rm crit}^2$ as a function of mass.  In the framework of
the wind-braking law, this can be achieved by scaling the convective
overturn time (Eq.\,\ref{eq:omegacrit}). We emphasize, however, that
it is not necessarily the overturn time itself that governs the
braking amplitude -- it is rather the resulting effect on the net
braking which is different (the value of $K$ may change as well). In
the left panel of Fig.\,\ref{fig:vsiniall}, we show such a scenario
for rotational braking in ultracool dwarfs. Solid lines indicate
cooling sequences for objects of 0.07--0.10\,M$_\sun$.  Dashed lines
mark isochrones at 1 and 10\,Gyr. The right panel in
Fig.\,\ref{fig:vsiniall} shows our choice of $K \omega_{\rm crit}^2$
as a function of mass. Asterisks show the values employed to calculate
the evolutionary tracks in the left panel of Fig.\,\ref{fig:vsiniall}.
Grey points connected by a straight line indicate a scaling according
to $\tau \propto M^{-2/3}$. Our scaling of $K \omega_{\rm crit}^2$
follows the relation

\begin{equation}
    K\omega_{\rm crit}^2 = -0.5 + 7.5 \frac{M}{M_\sun}.
\end{equation}

The braking law used in Fig.\,\ref{fig:vsiniall} provides a remarkable
similarity to the observed situation. Young low mass objects start at
intermediate rotation rates and accelerate during the first few tens
of millions of years.  Then, mass dependent wind braking causes
deceleration until they either lose most of their angular momentum at
an age of several ten million years (stars more massive than
0.09\,M$_\sun$), or the braking is so weak that until today even the
oldest objects have not lost all their angular momentum (objects less
massive than 0.09\,M$_\sun$). From the evolutionary tracks, one can
also predict ages and mass of the objects at least in a statistical
sense.  Without going into further detail we note that for the masses
of the three members of LHS~1070 (stars in Fig.\,\ref{fig:vsiniall})
relatively accurate predictions are achieved; masses from the orbital
solution are 0.12\,M$_{\sun}$ for component A (M5.5) and
0.079\,M$_{\sun}$ for components B and C \citep{Seifahrt07}.

In evaluating our braking law, we have not included the observations
of $v\,\sin{i}$ in T dwarfs (triangles in Fig.\,\ref{fig:vsiniall}).
They appear to rotate somewhat more slowly than the late-L dwarfs.
This can either be due to a different distribution of initial rotation
velocities (i.e.  differences during their formation mechanism), or
due to stronger braking reappearing in the mass regime below
50--60\,M$_{\rm Jup}$. As mentioned above, we are also skeptical about
a comparison between our data and the T dwarf velocities due to
systematic differences in the data (resolution) and their analysis.

\begin{figure*}
  \mbox{
    \includegraphics[width=.69\hsize,bbllx=0,bblly=0,bburx=630,bbury=335]{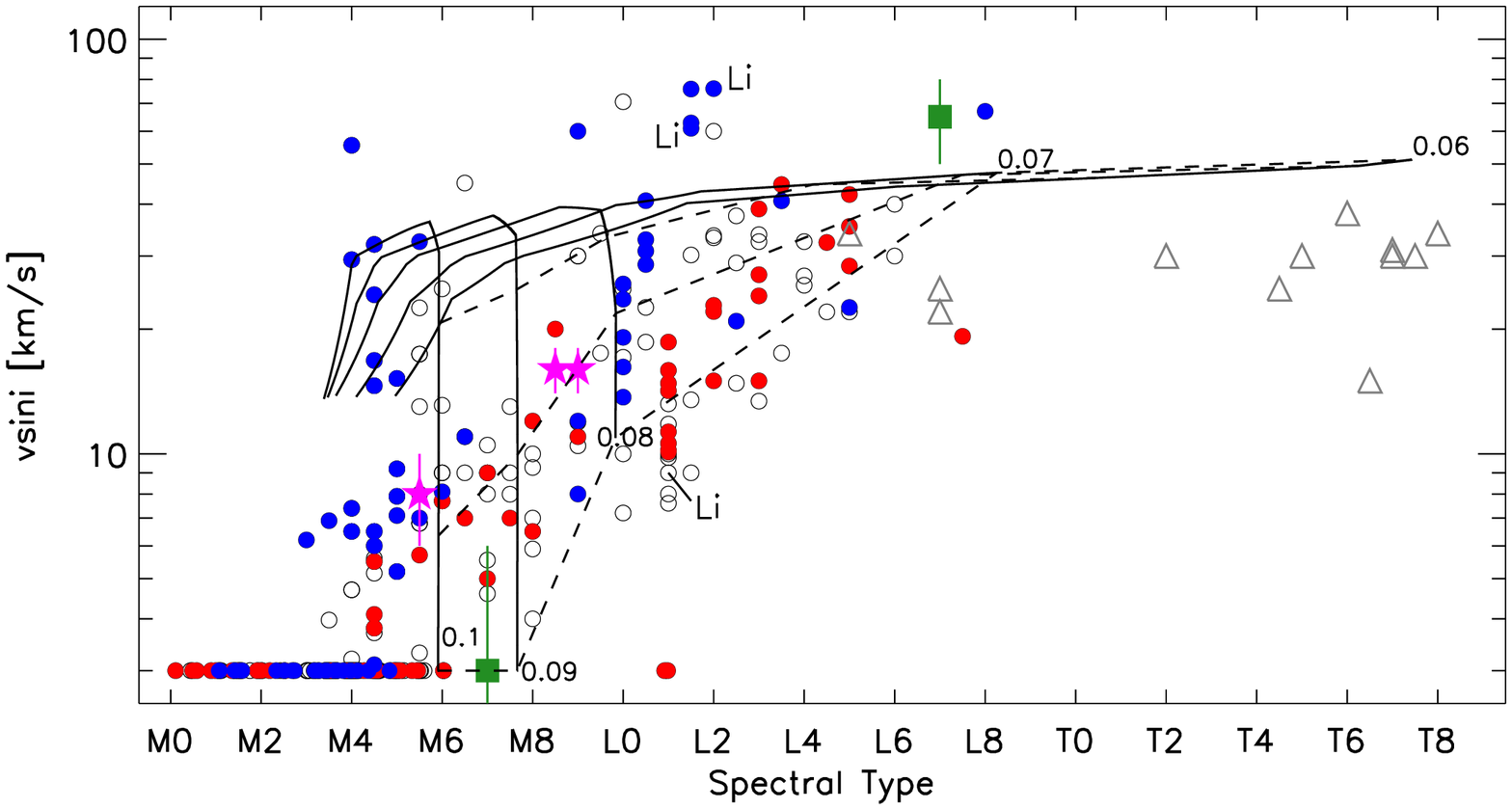}
    \includegraphics[width=.28\hsize,bbllx=100,bblly=-20,bburx=460,bbury=450]{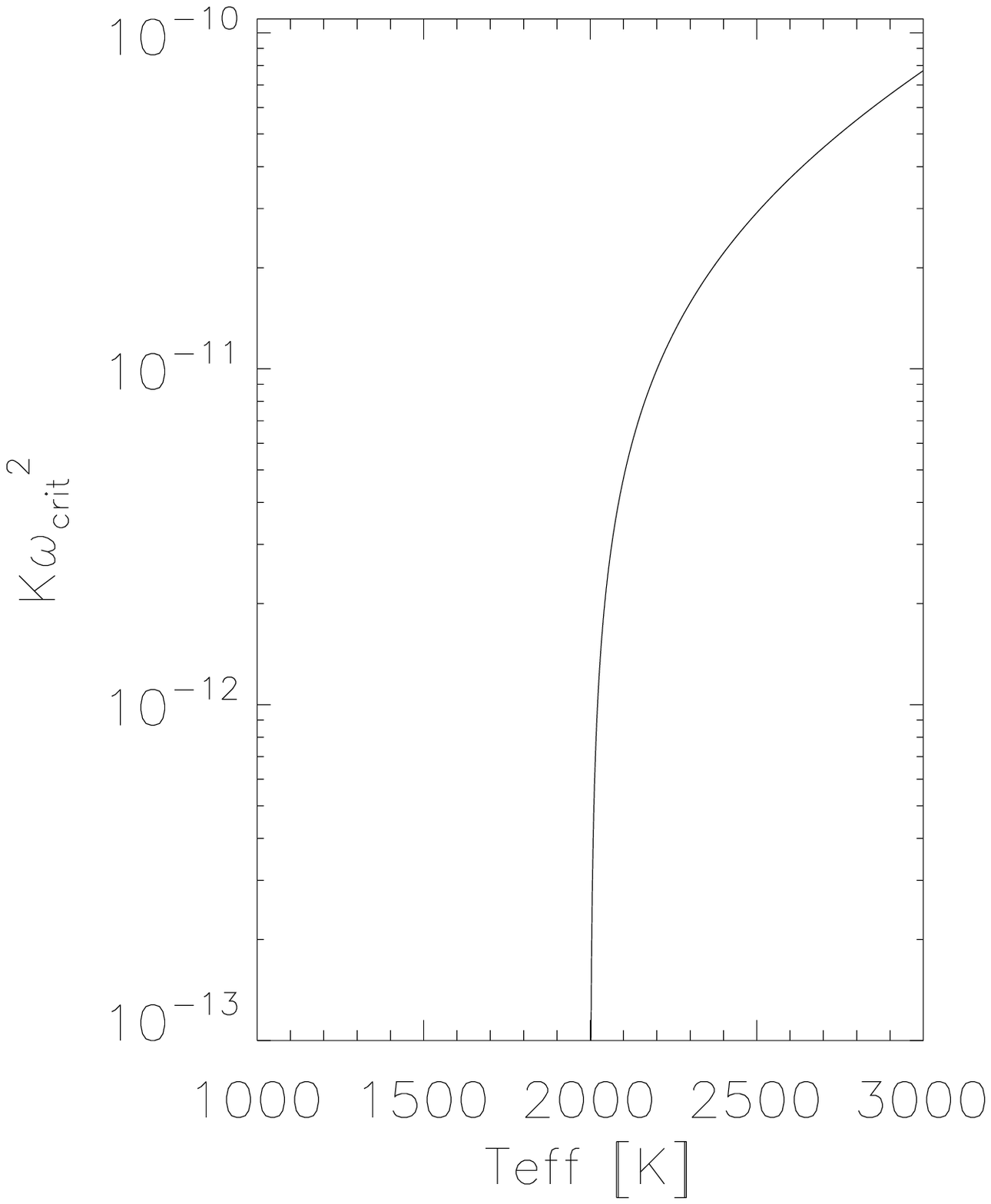}
  }
  \caption{\label{fig:vsiniall_teff}As Fig.\,\ref{fig:vsiniall} but
    with a braking law that scales with temperature.}
\end{figure*}

So far, we have described a braking law dependent on mass, which is
the most fundamental stellar parameter.  Atmospheric properties,
however, are not necessarily connected to mass, but are more closely
connected to temperature (and gravity). This is especially important
in brown dwarfs, since effective temperature changes a lot during
their evolution (while mass is a constant). Therefore, we consider the
effect of choosing $K \omega_{\rm crit}^2$ as a function of
temperature.  We searched for such a scaling law to see whether it
could explain the observed distribution of rotational velocities. The
best match we found is shown in Fig.\,\ref{fig:vsiniall_teff};
evolutionary tracks integrated over $T_{\rm eff}$ are shown in the
left panel, and $K \omega_{\rm crit}^2(T_{\rm eff})$ is shown in the
right panel. This model also does a good job of reproducing the lower
envelope of rotational velocities, and it does a slightly better job
with the extrapolation to the T dwarfs (but see above).  Such a
braking law has an easy physical explanation in the growing
atmospheric neutrality at very cool temperatures, and its effect on
coronal heating through diminished ability of the atmosphere to
produce magnetic dissipation \citep{MBS02}. A scatter (we assumed one
constant value) in the initial distribution of rotational velocities
could generate the high rotation rates observed in some young objects.
If the slower velocities in the T dwarfs were confirmed, slightly more
efficient braking at cool temperatures would be needed.

The main result of this section is that rotational evolution according
to a wind-braking law that scales with temperature can reproduce the
main features of rotational braking in L dwarfs. A very similar result
can be achieved with a law that depends on mass (through convective
overturn time as a proxy, with unclear physical meaning). In both
cases, very low-mass, cool objects must brake less efficiently.  A
very interesting consequence of both models is that the lower envelope
of minimum rotation velocities is directly coupled to the age of the
Galaxy: the objects simply haven't had enough time to decelerate any
further.

\subsection{The Connection between Rotation and Activity}

\begin{figure}
  \plotone{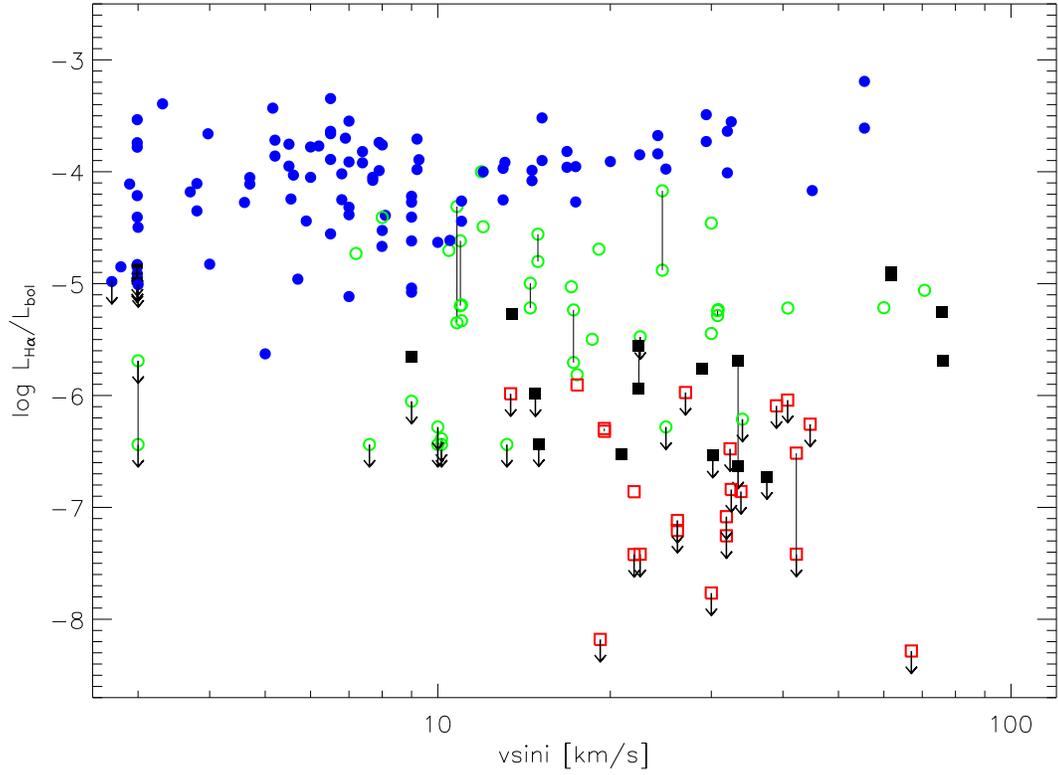}
  \caption{\label{fig:Lalpha_vsini}Normalized H$\alpha$ activity as a
    function of $v\,\sin{i}$. Solid lines and downward arrows are used
    as in Fig.\,\ref{fig:Halpha}. Filled blue circles show objects of
    spectral type earlier than M9; open green circles mark objects in
    the spectral type range M9--L1; filled black squares in the range
    L1.5--L3, and open red squares are used for objects L3.5 and
    later.}
\end{figure}

A breakdown of the rotation-activity connection known among solar-type
stars was noted at spectral type M9 by \citet{Mohanty03}.  They find a
saturation-type rotation-activity connection among M dwarfs cooler
than spectral type M4, and they report that such a relation no longer
exists in their sample of L dwarfs. We can now verify this conclusion
in our larger sample. In Fig.\,\ref{fig:Lalpha_vsini}, we show
normalized H$\alpha$ luminosity as a function of $v\,\sin{i}$ for our
sample. We use different symbols for different temperatures.  Multiple
observations of the same star are connected with solid lines and upper
limits in $L_{\rm{H_{\alpha}}}/L_{\rm{bol}}$ are marked with downward
arrows. A potential rotation-activity connection would best be visible
among objects of similar spectral type. That means looking, for
example, for open circles arranged in a pattern.

The well known rotation-activity relation among the M dwarfs (filled
circles) shows a $\Gamma$-shaped behavior: A large scatter appears in
$L_{\rm{H_{\alpha}}}/L_{\rm{bol}}$ in stars below the rotational
detection limit, while among the rapid rotators normalized H$\alpha$
luminosity is always very high.  Among the L dwarf subsamples,
however, we do not observe anything similar.  The three L dwarf
subgroups are different in their mean level of normalized H$\alpha$
luminosity, but they do not show any dependence on $v\,\sin{i}$. We
find no rotation-activity relation among the L dwarfs, in particular
we do not see an obvious connection between rotation and activity
within the plotted spectral type subgroups.  One possibility is that
since all the L dwarfs are rapidly rotating, they are all in the
rotationally saturated regime for magnetic field, while the
atmospheric heating decreases with temperature due to the neutrality
effect.

\begin{figure}
  \plottwo{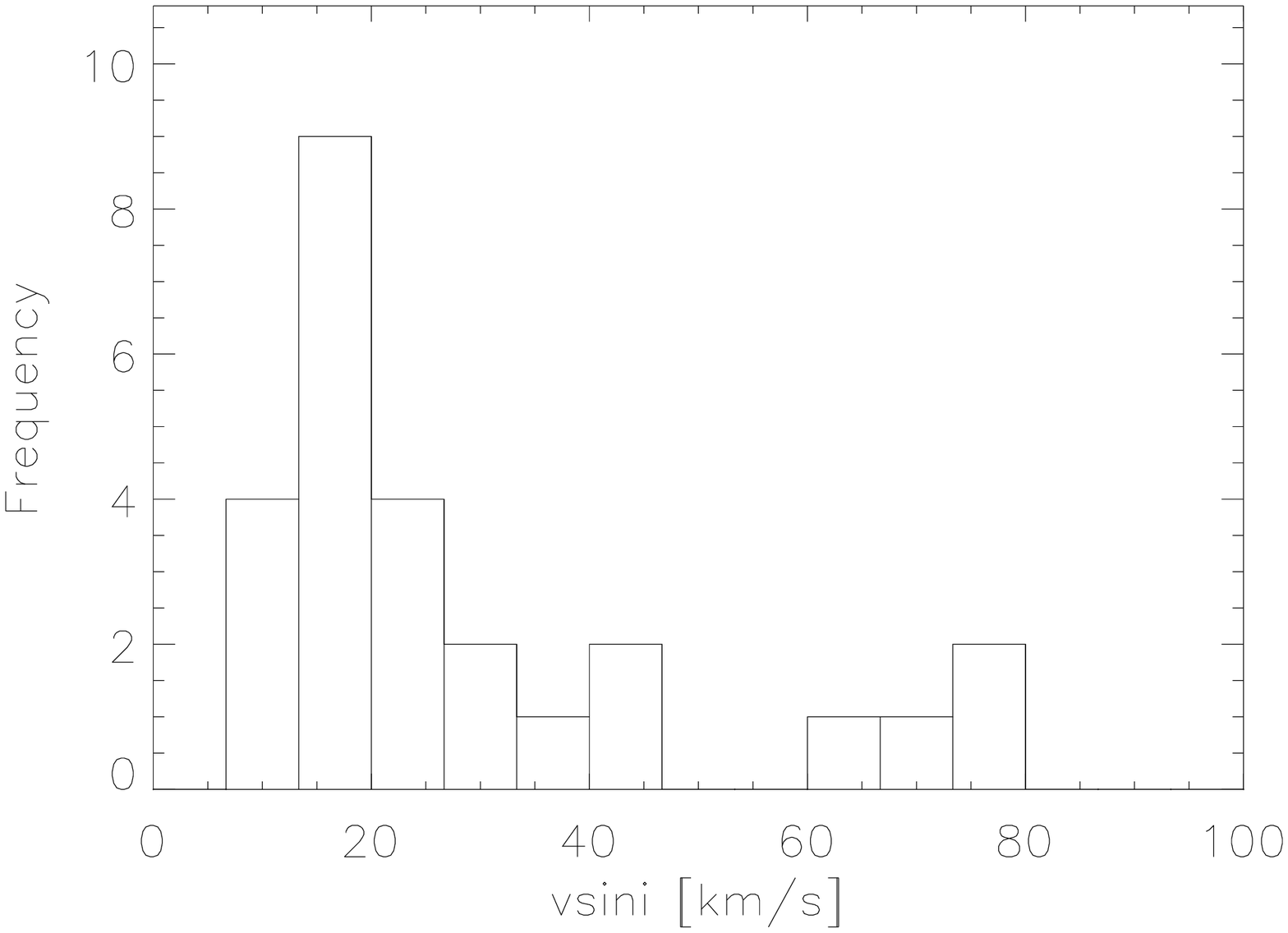}{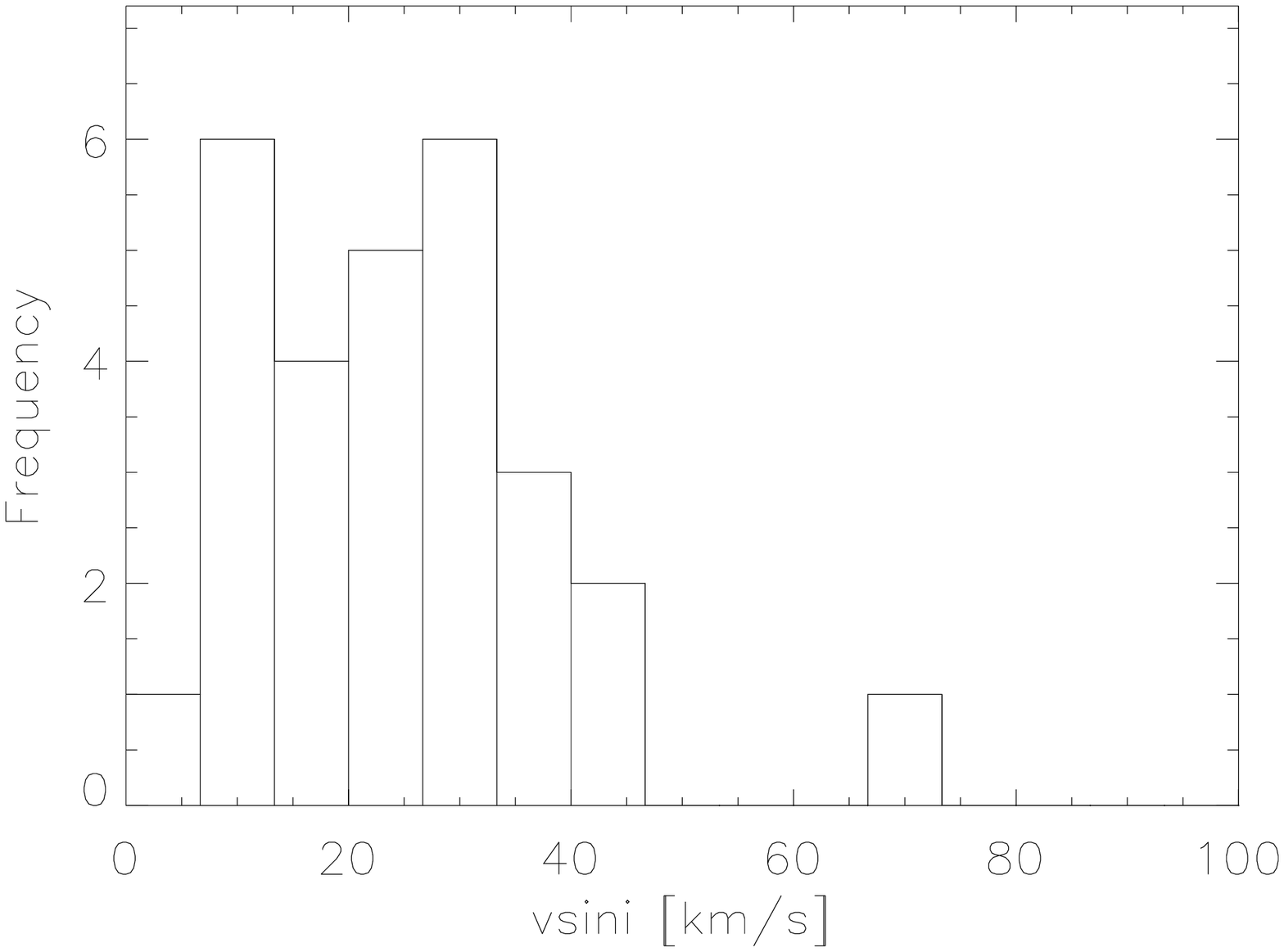}
  \caption{\label{fig:histo_vsini}Histogram of active (left panel) and
    inactive L dwarfs (right panel) in vsini.}
\end{figure}

An alternative way to search for a connection between rotation and
activity is to compare the distributions of active and inactive stars
as a function of rotation velocity. This is done in
Fig.\,\ref{fig:histo_vsini} where we plot the $v\,\sin{i}$
distribution among L dwarfs with detected H$\alpha$ emission and
without H$\alpha$ detection in the left and right panels,
respectively. We do not see an obvious difference between the two
distributions. A $\chi^2$-test shows that the probability to draw two
samples from one and the same underlying distribution and finding
differences larger than between the two observed here is 50\,\%.
Thus, there is no indication for a difference between the two
sub-samples. We conclude that rotation velocity is probably no longer
connected to activity observed in H$\alpha$ among the L dwarfs.

It is important to note that if L dwarfs are rotationally braked
during their lifetime, but activity does not scale with rotation, this
implies that no correlation between H$\alpha$ and age can be expected
due to rotational braking.

\section{Summary}

We have analyzed high resolution optical spectra of a comprehensive L
dwarf sample. From 61 spectra of 45 objects we measured the projected
rotation velocity $v\,\sin{i}$ and H$\alpha$ emission as an indicator
for chromospheric activity.

Half of our sample objects show activity in at least one spectrum.
The logarithmic normalized H$\alpha$ emission, $\log{L_{{\rm H}
    \alpha} / L_{\rm{bol}}}$, follows a smooth trend with lower
activity at later spectral types. The decrease of chromospheric
activity sets in around spectral type M5 \citep[see also][]{West04};
it connects smoothly to the average value of $\log{L_{{\rm H} \alpha}
  / L_{\rm{bol}}} \approx -4$ in early M dwarfs. The fraction of L
dwarfs with detected H$\alpha$ emission also decreases towards later
spectral types. This is consistent with the results from
\citet{West04} and \citet{Schmidt07}, but our fraction is somewhat
higher than theirs, probably because of the higher sensitivity of our
data.  Activity is observed in objects as late as L5. We did not find
H$\alpha$ emission in the 2 L dwarfs of very late spectral type.

For 15 objects we possess more than one spectrum; 8 of them show
variable H$\alpha$ emission. We consider this variability as a sign of
robust flaring activity among L dwarfs. This means that flaring
activity is a common process among (early) L dwarfs. The fraction of
variable objects ($53\pm13\,\%$) among objects with more than one
observation is almost identical to the fraction of active objects
($51\pm7\,\%$) in the full sample. Our data is consistent with the
assumption that all active L dwarfs are variable.

Our rotation velocities add substantially to the available rotation
measurements in brown dwarfs. Between spectral types M7 and L8,
objects of later spectral type rotate faster than earlier-type
objects. Minimum (projected) rotation velocities are about
5\,km\,s$^{-1}$ at spectral type L0 and about 20--30\,km\,s$^{-1}$ at
spectral type L6. One L1 dwarf shows no measurable rotation, which we
attribute to low inclination. This object allows us to measure
rotations using a template spectrum of a non-rotating L dwarf.

We have excluded the possibility that the lack of slowly rotating
rotating L dwarfs is an observational bias by investigating the
age-distribution of the objects using primarily tangential velocities.
We suggest that the distribution of rotation velocities is probably
induced by mass-dependent rotational braking with weaker braking at
lower temperatures. We find remarkably good agreement between the
observed distribution of rotational velocities and predictions from a
wind-braking model similar to models applied to higher mass stars
(with the caveat that high initial angular momentum is required). In
our model, braking scales with temperature, perhaps due to decreasing
magnetic dissipation, although another mass-dependent law can work as
well.

Another physical explanation that has been offered is the possibly
changing magnetic topology in low mass objects.  \citet{Durney93}
suggested that fully convective dynamos might produce predominantly
small scale fields, leading to weaker rotational braking than
large-scale fields generated in sun-like stars can produce. This
suggestion has now run afoul of both observations \citep{Donati06} and
theory \citep{Dobler06, Browning08}. It does seem that something
happens in angular momentum evolution near the fully convective
boundary; that is where measurable Doppler rotation begins to reappear
in a reasonable fraction of stars. The question is whether that is
coincidental or not. The real effects of temperature-induced
neutrality don't kick in until several spectral sub-classes later, so
it may be that there is more to the full picture than that.

We have searched for a rotation-activity connection among L dwarfs.
While normalized H$\alpha$ activity is decreasing with later spectral
type, no trend with rotation is visible among subgroups of comparable
spectral types. One reasonable explanation is that the field
generation may be saturated because of the rapid rotations, or the
conversion of fields to excess emission is different or more unstable.
It is not yet clear how magnetic fields behave on L dwarfs, nor how
they are converted into H$\alpha$, X-ray, or radio-emission excesses.
This should be a fertile area for research in the next few years.


\acknowledgments

We thank an anonymous referee for thorough reading and very helpful
comments.  This work is based on observations obtained from the
European Southern Observatory, PIDs 077.C-0449 and 078.C-0025, and on
observations obtained from the W.M.  Keck Observatory, which is
operated as a scientific partnership among the California Institute of
Technology, the University of California and the National Aeronautics
and Space Administration. We would like to acknowledge the great
cultural significance of Mauna Kea for native Hawaiians and express
our gratitude for permission to observe from atop this mountain. Based
on observations made with the European Southern Observatory telescopes
obtained from the ESO/ST-ECF Science Archive Facility. A.R.
acknowledges research funding from the DFG as an Emmy Noether fellow
(RE 1664/4-1). G.B. acknowledges support from the NSF through grant
AST-0606748.







\begin{deluxetable}{rcccr}
  \tablecaption{\label{tab:observations} Table of observations}
  \tablewidth{0pt}
  \tablehead{\colhead{Name} & \colhead{SpType} & \colhead{$J$} & \colhead{exp. time} & \colhead{S/N}\\
    &&[mag] & [s]} 
  \startdata 
  \cutinhead{Keck observations}
  2MASS J$12212770+0257198$ & L0.0 & 13.17  & 2700 & 20 \\
  2MASS J$17312974+2721233$ & L0.0 & 12.09  & 1200 & 20 \\
  2MASS J$18544597+8429470$ & L0.0 & 13.66  & 4200 & 15 \\
  2MASS J$07464256+2000321$ & L0.5 & 11.76  & 1800 & 25 \\
  2MASS J$23515044-2537367$ & L0.5 & 12.47  & 3600 & 30 \\
  2MASS J$06023045+3910592$ & L1.0 & 12.30  & 3000 & 15 \\
  2MASS J$10224821+5825453$ & L1.0 & 13.50  & 3600 & 15 \\
  2MASS J$10224821+5825453$ & L1.0 & 13.50  & 1800 & 15 \\
  2MASS J$10452400-0149576$ & L1.0 & 13.16  & 7200 & 20 \\
  2MASS J$10484281+0111580$ & L1.0 & 12.92  & 1800 & 20 \\
  2MASS J$13004255+1912354$ & L1.0 & 12.72  & 1800 & 25 \\
  2MASS J$14392836+1929149$ & L1.0 & 12.76  & 2000 & 25 \\
  2MASS J$15551573-0956055$ & L1.0 & 12.56  & 2100 & 15 \\
  2MASS J$18071593+5015316$ & L1.5 & 12.93  & 5100 & 45 \\
  2MASS J$20575409-0252302$ & L1.5 & 13.12  & 3600 & 25 \\
  2MASS J$08283419-1309198$ & L2.0 & 12.80  & 1200 & 20 \\
  2MASS J$11553952-3727350$ & L2.0 & 12.81  & 1800 & 15 \\
  2MASS J$13054019-2541059$ & L2.0 & 13.41  & 5700 & 30 \\
  2MASS J$15065441+1321060$ & L3.0 & 13.37  & 3600 & 20 \\
  2MASS J$21041491-1037369$ & L3.0 & 13.84  & 3600 & 15 \\
  2MASS J$00361617+1821104$ & L3.5 & 12.47  & 6000 & 20 \\
  2MASS J$07003664+3157266$ & L3.5 & 12.92  & 1500 & 20 \\
  \cutinhead{UVES observations}
  2MASS J$03140344+1603056$ & L0.0 & 12.53  & 4000 & 15 \\
  2MASS J$12212770+0257198$ & L0.0 & 13.17  & 5400 & 20 \\
  2MASS J$17312974+2721233$ & L0.0 & 12.09  & 2400 & 25 \\
  2MASS J$22000201-3038327$ & L0.0 & 14.36  & 6900 & 20 \\
  2MASS J$07464256+2000321$ & L0.5 & 11.76  & 1500 & 20 \\
  2MASS J$02355993-2331205$ & L1.0 & 13.67\tablenotemark{$\star$}  & 3600 &  9 \\
  2MASS J$10452400-0149576$ & L1.0 & 13.16  & 5400 & 15 \\
  2MASS J$10484281+0111580$ & L1.0 & 12.92  & 4200 & 15 \\
  2MASS J$13004255+1912354$ & L1.0 & 12.72  & 3900 & 20 \\
  2MASS J$13595510-4034582$ & L1.0 & 13.65  & 5640 & 25 \\
  2MASS J$15551573-0956055$ & L1.0 & 12.56  & 3000 & 20 \\
  2MASS J$16452211-1319516$ & L1.5 & 12.45  & 2400 & 20 \\
  2MASS J$20575409-0252302$ & L1.5 & 13.12  & 5100 & 20 \\
  2MASS J$08283419-1309198$ & L2.0 & 12.80  & 3600 & 15 \\
  2MASS J$09211410-2104446$ & L2.0 & 12.78  & 3600 & 20 \\
  2MASS J$11553952-3727350$ & L2.0 & 12.81  & 4200 & 25 \\
  2MASS J$05233822-1403022$ & L2.5 & 13.08  & 4800 & 20 \\
  2MASS J$15065441+1321060$ & L3.0 & 13.37  & 6600 & 30 \\
  2MASS J$17054834-0516462$ & L4.0 & 13.31  & 6300 & 10 \\
  2MASS J$17054834-0516462$ & L4.0 & 13.31  & 6300 & 20 \\
  2MASS J$00043484-4044058$ & L5.0 & 13.11  & 5100 & 10 \\
  2MASS J$08354256-0819237$ & L5.0 & 13.17  & 5700 & 15 \\
  2MASS J$15074769-1627386$ & L5.0 & 12.83  & 4200 & 30 \\
  2MASS J$02550357-4700509$ & L8.0 & 13.25  & 6000 & 20 \\
  \cutinhead{UVES archive data}
  2MASS J$11593850+0057268$ & L0.0 & 14.08  & 2700 & 15 \\
  2MASS J$07464256+2000321$ & L0.5 & 11.76  & \phn900 & 25 \\
  2MASS J$14122449+1633115$ & L0.5 & 13.89  & 2700 & 10 \\
  2MASS J$14413716-0945590$ & L0.5 & 14.02  & 3600 & 15 \\
  2MASS J$14392836+1929149$ & L1.0 & 12.76  & \phn900 & 15 \\
  2MASS J$11455714+2317297$ & L1.5 & 15.39  & 7200 &  6 \\
  2MASS J$13340623+1940351$ & L1.5 & 15.47  & 7200 &  7 \\
  2MASS J$10292165+1626526$ & L2.5 & 14.29  & 3600 & 10 \\
  2MASS J$10473109-1815574$ & L2.5 & 14.12  & 7200 & 10 \\
  2MASS J$09130320+1841501$ & L3.0 & 15.97  & 7200 &  5 \\
  2MASS J$12035812+0015500$ & L3.0 & 14.01  & 9000 & 10 \\
  2MASS J$16154416+3559005$ & L3.0 & 14.54  & 7200 & 15 \\
  2MASS J$22244381-0158521$ & L4.5 & 14.07  & 4200 & 10 \\
  2MASS J$15074769-1627386$ & L5.0 & 12.83  & 1800 & 20 \\
  2MASS J$08251968+2115521$ & L7.5 & 15.10  & 9000 &  6 \\
  \enddata
  \tablenotetext{$\star$}{\citet{Gizis01}}
\end{deluxetable}

\begin{deluxetable}{cl}
  \tablecaption{\label{tab:chifactor}Parameters of the fifth order
    polynomial fit to $\log{\chi} (T_{\rm{eff}})$ in
    Eq.\,\ref{eq:chifactor}} \tablehead{\colhead{Parameter} &
    \colhead{Value}} \startdata
  a & $-   6.73424e{+1}$\\
  b & \phs$1.11938e{-1}$\\
  c & $-   8.26212e{-5}$\\
  d & \phs$3.04492e{-8}$\\
  e & $-   5.51137e{-12}$\\
  f & \phs$3.90255e{-16}$
  \enddata
\end{deluxetable}

\begin{deluxetable}{ccrc}
  \tablecaption{\label{tab:halpha} Targets with variable H$\alpha$
    emission} \tablewidth{0pt} \tablehead{\colhead{Name} &
    \colhead{Date of observation} & \colhead{H$\alpha$ EqW} &
    \colhead{$\log{\frac{L_{\rm H{\alpha}}}{L_{\rm bol}}}$}}
  \startdata
  2MASS J$1221+02$ & 2006-05-12 & 25.65& \phantom{$<$}$-4.18$\\
  (L0.0)       & 2006-04-10 &  5.01& \phantom{$<$}$-4.88$\\[1mm]
  2MASS J$1731+27$ & 2005-05-13 & 10.51& \phantom{$<$}$-4.57$\\
  (L0.0)       & 2006-04-15 &  5.99& \phantom{$<$}$-4.80$\\[1mm]
  2MASS J$1022+58$ & 2006-05-14 &  3.32& \phantom{$<$}$-5.22$\\
  (L1.0)       & 2006-05-19 &  5.52& \phantom{$<$}$-5.00$\\[1mm]
  2MASS $1048+01$  & 2006-05-12 &  3.19& \phantom{$<$}$-5.23$\\
  (L1.0)       & 2007-01-30 &  1.08& \phantom{$<$}$-5.71$\\[1mm]
  2MASS J$1439+19$ & 2000-04-27 &  3.48& \phantom{$<$}$-5.20$\\
  (L1.0)       & 2005-03-02 & 13.22& \phantom{$<$}$-4.62$\\[1mm]
  2MASS J$1555-09$ & 2006-04-02 &  2.45& \phantom{$<$}$-5.35$\\
  (L1.0)       & 2006-05-13 & 26.79& \phantom{$<$}$-4.31$\\[1mm]
  2MASS J$0828-13$ & 2005-12-17 &  1.75& \phantom{$<$}$-5.68$\\
  (L2.0)       & 2006-11-08 &$<$ 0.20& $<-6.63$\\[1mm]
  2MASS J$1155-37$ & 2006-05-12 &  1.00& \phantom{$<$}$-5.96$\\
  (L2.0)       & 2006-04-09 &  2.39& \phantom{$<$}$-5.54$\\[1mm]
  \enddata
\end{deluxetable}

\begin{deluxetable}{rccrrrr}
  \tablecaption{\label{tab:results} Results from our analysis.
    Effective temperature is from the calibration of
    \citet{Golimowski04}, $v_{\rm{tan}}$ is from \cite{Schmidt07}.
    Objects marked with a $\star$ were observed more than once. For
    them, we give the values of lowest activity in the table.}
  \tablewidth{0pt}
  \tablehead{\colhead{Name} & \colhead{SpType} & \colhead{$T_{\rm eff}$} & \colhead{$v\,\sin{i}$} & \colhead{H$\alpha$ EqW} & \colhead{log($\frac{L_\mathrm{H\alpha}}{L_\mathrm{bol}}$)} & \colhead{$v_{\rm{tan}}$}\\
    && [K] & [km\,s$^{-1}$] &[\AA]&& [km\,s$^{-1}$]} \startdata
  2MASS $0314+16$ & L0.0 &  2300 &   19 &   $   7.72$ &   $  -4.69$\phantom{$^*$} & 16.8 \\
2MASS $1159+00$ & L0.0 &  2300 &   71 &   $   3.31$ &   $  -5.06$\phantom{$^*$} &      \\
2MASS $1221+02$ & L0.0 &  2300 &   25 &   $   5.01$ &   $  -4.88^*$ &  8.1 \\
2MASS $1731+27$ & L0.0 &  2300 &   15 &   $   5.99$ &   $  -4.80^*$ & 15.8 \\
2MASS $1854+84$ & L0.0 &  2300 &    7 &   $   7.06$ &   $  -4.73$\phantom{$^*$} &      \\
2MASS $2200-30$ & L0.0 &  2300 &   17 &   $   3.56$ &   $  -5.03$\phantom{$^*$} &      \\
2MASS $0746+20$ & L0.5 &  2250 &   31 &   $   2.36$ &   $  -5.29^*$ & 21.3 \\
2MASS $1412+16$ & L0.5 &  2250 &   19 &   $   1.45$ &   $  -5.50$\phantom{$^*$} &      \\
2MASS $1441-09$ & L0.5 &  2250 &   23 &   $<  1.53$ &   $< -5.48$\phantom{$^*$} &      \\
2MASS $2351-25$ & L0.5 &  2250 &   41 &   $   2.76$ &   $  -5.22$\phantom{$^*$} & 26.4 \\
2MASS $0235-23$ & L1.0 &  2200 &   13 &   $<  0.20$ &   $< -6.44$\phantom{$^*$} &      \\
2MASS $0602+39$ & L1.0 &  2200 &    9 &   $<  0.49$ &   $< -6.05$\phantom{$^*$} &      \\
2MASS $1022+58$ & L1.0 &  2200 &   15 &   $   3.32$ &   $  -5.22^*$ & 81.9 \\
2MASS $1045-01$ & L1.0 &  2200 &$<$ 3 &   $<  0.20$ &   $< -6.44^*$ & 36.6 \\
2MASS $1048+01$ & L1.0 &  2200 &   17 &   $   1.08$ &   $  -5.71^*$ & 37.9 \\
2MASS $1300+19$ & L1.0 &  2200 &   10 &   $<  0.20$ &   $< -6.44^*$ & 97.9 \\
2MASS $1359-40$ & L1.0 &  2200 &    8 &   $<  0.20$ &   $< -6.44$\phantom{$^*$} &      \\
2MASS $1439+19$ & L1.0 &  2200 &   11 &   $   3.48$ &   $  -5.20^*$ & 89.0 \\
2MASS $1555-09$ & L1.0 &  2200 &   11 &   $   2.45$ &   $  -5.35^*$ &      \\
2MASS $1145+23$ & L1.5 &  2140 &   14 &   $   3.69$ &   $  -5.27$\phantom{$^*$} &      \\
2MASS $1334+19$ & L1.5 &  2140 &   30 &   $<  0.20$ &   $< -6.53$\phantom{$^*$} &      \\
2MASS $1645-13$ & L1.5 &  2140 &    9 &   $   1.51$ &   $  -5.66$\phantom{$^*$} &      \\
2MASS $1807+50$ & L1.5 &  2140 &   76 &   $   3.79$ &   $  -5.26$\phantom{$^*$} & 11.2 \\
2MASS $2057-02$ & L1.5 &  2140 &   62 &   $   8.15$ &   $  -4.92^*$ &  5.9 \\
2MASS $0828-13$ & L2.0 &  2080 &   33 &   $<  0.20$ &   $< -6.63^*$ &      \\
2MASS $0921-21$ & L2.0 &  2080 &   15 &   $<  0.32$ &   $< -6.42$\phantom{$^*$} & 57.8 \\
2MASS $1155-37$ & L2.0 &  2080 &   22 &   $   1.00$ &   $  -5.96^*$ & 49.3 \\
2MASS $1305-25$ & L2.0 &  2080 &   76 &   $   1.75$ &   $  -5.69$\phantom{$^*$} & 25.7 \\
2MASS $0523-14$ & L2.5 &  2010 &   21 &   $   0.34$ &   $  -6.52$\phantom{$^*$} &  5.1 \\
2MASS $1029+16$ & L2.5 &  2010 &   29 &   $   1.96$ &   $  -5.76$\phantom{$^*$} &      \\
2MASS $1047-18$ & L2.5 &  2010 &   15 &   $<  1.16$ &   $< -5.99$\phantom{$^*$} &      \\
2MASS $0913+18$ & L3.0 &  1950 &   34 &   $<  0.20$ &   $< -6.86$\phantom{$^*$} &      \\
2MASS $1203+00$ & L3.0 &  1950 &   39 &   $<  1.17$ &   $< -6.09$\phantom{$^*$} & 88.1 \\
2MASS $1506+13$ & L3.0 &  1950 &   20 &   $   0.69$ &   $  -6.32^*$ & 74.2 \\
2MASS $1615+35$ & L3.0 &  1950 &   13 &   $<  1.50$ &   $< -5.98$\phantom{$^*$} &      \\
2MASS $2104-10$ & L3.0 &  1950 &   27 &   $<  1.53$ &   $< -5.97$\phantom{$^*$} & 58.9 \\
2MASS $0036+18$ & L3.5 &  1880 &   45 &   $<  1.09$ &   $< -6.26$\phantom{$^*$} & 36.9 \\
2MASS $0700+31$ & L3.5 &  1880 &   41 &   $<  1.78$ &   $< -6.04$\phantom{$^*$} & 29.7 \\
2MASS $1705-05$ & L4.0 &  1820 &   26 &   $<  0.20$ &   $< -7.12^*$ &      \\
2MASS $2224-01$ & L4.5 &  1760 &   32 &   $<  1.21$ &   $< -6.48$\phantom{$^*$} & 52.8 \\
2MASS $0004-40$ & L5.0 &  1700 &   42 &   $<  0.20$ &   $< -7.42$\phantom{$^*$} & 83.3 \\
2MASS $0835-08$ & L5.0 &  1700 &   23 &   $<  0.20$ &   $< -7.42$\phantom{$^*$} & 24.0 \\
2MASS $1507-16$ & L5.0 &  1700 &   32 &   $<  0.29$ &   $< -7.25^*$ & 30.8 \\
2MASS $0825+21$ & L7.5 &  1500 &   19 &   $<  0.20$ &   $< -8.18$\phantom{$^*$} & 38.2 \\
2MASS $0255-47$ & L8.0 &  1480 &   67 &   $<  0.20$ &   $< -8.28$\phantom{$^*$} & 28.6 \\

  \enddata
\end{deluxetable}

\clearpage

\newpage

\end{document}